\documentclass[11pt,journal,onecolumn,draftclsnofoot]{IEEEtran}
\ifCLASSINFOpdf
\else
\fi

\usepackage[utf8]{inputenc} 
\usepackage[bookmarks, colorlinks=true, plainpages = false, citecolor = blue,linkcolor=red,urlcolor = blue, filecolor = blue,pagebackref]{hyperref}

\usepackage[T1]{fontenc}    
\usepackage{url}            
\usepackage{booktabs}       
\usepackage{amssymb}       
\usepackage{amsthm}
\usepackage{nicefrac}       
\usepackage{microtype}      
\usepackage{graphicx}
\usepackage{nicefrac}
\usepackage{amscd, MnSymbol,mathrsfs}
\usepackage{bm}
\DeclareMathAlphabet{\mathbbm}{U}{bbm}{m}{n}
\usepackage[caption=false,font=normalsize,labelfont=sf,textfont=sf]{subfig}
\usepackage{algorithmicx}
\usepackage[linesnumberedhidden,ruled,vlined]{algorithm2e}
\usepackage{epstopdf}
\usepackage{color}

\usepackage{cite}

\usepackage{amsthm}\usepackage{dsfont}\usepackage{array}\usepackage{mathrsfs}\usepackage{comment}

\theoremstyle{plain}\newtheorem{theorem}{\textbf{Theorem}}\newtheorem{assumption}{\textbf{Assumption}}\newtheorem{proposition}{\textbf{Proposition}}

\theoremstyle{definition}

\theoremstyle{remark}

\newcommand{\balpha}{{\boldsymbol \alpha}}

\newcommand{\bA}{{\boldsymbol A}}

\newcommand{\be}{{\boldsymbol e}}

\newcommand{\bw}{{\boldsymbol w}}
\newcommand{\bx}{{\boldsymbol x}}
\newcommand{\by}{{\boldsymbol y}}

\newcommand{\E}{{\mathbb E}}
\newcommand{\R}{{\mathbb R}}

\newcommand{\argmax}{\mathrm{argmax}}

\begin{document}
%
\title{Distributed Dual Coordinate Ascent in General Tree Networks and Communication Network Effect on Synchronous Machine Learning}
%
%
%

\author{Myung~Cho,
        Lifeng~Lai,
        and~Weiyu~Xu
\thanks{M. Cho is with the Department
of ECE, Penn State Behrend, Erie, PA 16563, USA (E-mail: mxc6077@psu.edu).}
\thanks{L. Lai is with the Department of ECE, University of California, Davis, CA 95616, USA (E-mail: lflai@ucdavis.edu).} 
\thanks{W. Xu is with the Department of ECE, University of Iowa, Iowa City, IA 52242, USA (E-mail: weiyu-xu@uiowa.edu).}
}

\maketitle
\begin{abstract}
\textcolor[rgb]{0,0,0}{Due to the big size of data and limited data storage volume of a single computer or a single server, data are often stored in a distributed manner. Thus, performing large-scale machine learning operations with the distributed datasets through communication networks is often required. In this paper, we study the convergence rate of the distributed dual coordinate ascent for distributed machine learning problems in a general tree-structured network. Since a tree network model can be understood as the generalization of a star network model, our algorithm can be thought of as the generalization of the distributed dual coordinate ascent in a star network model. We provide the convergence rate of the distributed dual coordinate ascent over a general tree network in a recursive manner and analyze the network effect on the convergence rate. Secondly, by considering network communication delays, we optimize the distributed dual coordinate ascent algorithm to maximize its convergence speed. From our analytical result, we can choose the optimal number of local iterations depending on the communication delay severity to achieve the fastest convergence speed. In numerical experiments, we consider machine learning scenarios over communication networks, where local workers cannot directly reach to a central node due to constraints in communication, and demonstrate that the usability of our distributed dual coordinate ascent algorithm in tree networks. Additionally, we show that adapting number of local and global iterations to network communication delays in the distributed dual coordinated ascent algorithm can improve its convergence speed. } 
\end{abstract}


%
\IEEEpeerreviewmaketitle
\section{Introduction}
\label{sec:intro}
In the past decade, machine learning has been driven by huge amount of data, simply called \textit{big data}. In various fields including education, finance, transportation, healthcare, engineering, and management, etc., big data is fundamentally changing our lives and societies \cite{Chen2014Big}, e.g., recommender services \cite{verma2015big}, disease diagnosis and analysis \cite{andreu2015big}, or even signal recovery \cite{efromovich2004data}.  However, due to limited storage volumes in storage server and constraints in communication, we face challenges of processing big data. Especially, big data are very often collected and stored from different locations at different times. Also, it is very expensive, inefficient, and insecure to aggregate distributed data in one central place. \textcolor[rgb]{0,0,0}{Machine learning over wireless communication networks can be a good example having these challenges, where machine learning process is performed through multiple decentralized devices having local data over wireless communication networks without sharing their raw data with others \cite{park2019wireless,zhu2020toward}.} Therefore, it is quite natural to consider solving large-scale machine learning problems with distributed data over communication networks in order to obtain valuable information from the distributed data.

Solving large-scale machine learning problems dealing with distributed data over communication networks is a challenging problem, due to the limited resources and obstacles including limited communication bandwidth, limited storage volume, limited energy consumption or even privacy and security issues. In order to handle the challenges of distributed data with limited resources, researchers have developed and studied various algorithms in \cite{chen2016revisiting,gemulla2011large,yang2013trading,jaggi2014communication,ma2015adding,zhao2016fast,hsieh2008dual,shalev2013stochastic,zhang2015fast,huo2016distributed,hsieh2015PASSCoDe} and the references therein. More specially, synchronous Stochastic Gradient Decent (SGD) \cite{chen2016revisiting,gemulla2011large}, synchronous Stochastic Dual Coordinate Ascent (SDCA) \cite{hsieh2008dual,yang2013trading,jaggi2014communication,ma2015adding,shalev2013stochastic}, asynchronous SGD \cite{zhao2016fast,zhang2015fast}, and asynchronous SDCA \cite{huo2016distributed,hsieh2015PASSCoDe} for distributed data have been intensively investigated in the literature. Among them,  \cite{hsieh2008dual} reports that even though the convergence of SGD does not depend on the size of data, SDCA can outperform SGD when we need relatively high solution accuracy. Moreover, asynchronous updating scheme in SGD and SDCA can suffer from the conflicts between intermediate results. 

Motivated by these facts, \cite{yang2013trading,jaggi2014communication,ma2015adding} consider using synchronous SDCA to solve regularized loss minimization problems in a star network. In the scenario, data are distributed over a few local workers in the star network, and each local worker communicates with a central station. The authors in \cite{yang2013trading,jaggi2014communication,ma2015adding} analyze the convergence rate of the distributed SDCA in terms of communication rounds.  Espeically, the strong aspects of the proposed distributed optimization framework in \cite{jaggi2014communication,ma2015adding} include free-of-tuning parameters or learning rates compared with SGD-based methods, and the readily computable duality gap for fair stopping criterion and efficient accuracy certificates. 

However, in practice, the local workers may be organized in various types of network topologies such as a tree, a mesh, or a ring. \textcolor[rgb]{0,0,0}{Especially, in wireless communication networks, due to limited communication power and energy consumption, local workers sometimes cannot directly communicate with a central node. In this situation, the distributed dual coordinate ascent in a star network cannot be  used for distributed machine learning. And if intermediate nodes are added for the communication from local workers to a central node, the distributed dual coordinate ascent for a star network will easily suffer from the increased latency and delay in communication. Therefore, considering communication network topologies in distributed machine learning problems is an important problem, and taking advantage of the network topologies may play a significant role in finding efficient solutions for the problems.} Then, it is natural to ask how to design and analyze the distributed dual coordinate ascent over a network with general topologies beyond a star network. Additionally, since delay and latency in communication can affect the convergence speed of a distributed machine learning algorithm, it is essential to investigate how network communication delays will affect the design and convergence rate of distributed dual coordinate ascent algorithms previsouly introduced in \cite{yang2013trading,jaggi2014communication,ma2015adding} in terms of overall computational time instead of the number of communication rounds.  The authors in \cite{tsianos2012communication} analyzed the convergence bound in terms of time by considering communication delays in a network for a consensus optimization problem. Additionally, the research \cite{ying2019supervised,chang2015multi,shi2014linear,Mateos2010Distributed,Hong2017Stochastic} studied separable consensus problems to each worker by using ADMM techniques. We remark that the regularized loss minimization problem considered in \cite{yang2013trading,jaggi2014communication,ma2015adding} is a different problem from the consensus problems considered in \cite{ying2019supervised,chang2015multi,shi2014linear,Mateos2010Distributed,Hong2017Stochastic} in the aspect of separability. \textcolor[rgb]{0,0,0}{Moreover, in \cite{carreira2014distributed,zeng2019global,lau2018proximal}, the authors considered  a distributed deep neural network model. By introducing auxiliary variables, the authors made the non-convex problem separable, which can lead to a consensus problem over a network. Unlike the works in \cite{carreira2014distributed,zeng2019global,lau2018proximal}, we consider to solve distributed machine learning problems in an augmented manner by taking into account communication networks. Therefore, our work focuses on the communication and network topology's effect on the distributed machine learning algorithms, while the works in \cite{carreira2014distributed,zeng2019global,lau2018proximal} focus on the alternating or block coordinate descent algorithm itself without considering network constraints to solve neural network problems with auxiliary variables.}


\textcolor[rgb]{0,0,0}{The contribution of this paper is three-fold. Firstly, we design the distributed dual coordinate ascent for a regularized loss minimization problem in a general \emph{tree-structured} communication network and analyze the convergence rate of the algorithm over the general tree network. Since a star network is a special case of a general tree network, our distributed dual coordinate ascent algorithm can be thought of as a generalized version of the distributed dual coordinate ascent in a star network. Secondly, we study the influence of the communication constraints in a network on the convergence rate of the distributed dual coordinate ascent. By considering  delays in communication, we optimize the network-constrained dual coordinate ascent to maximize its convergence speed in terms of time, and provide an analytical solution for the optimal number of local iterations depending on the communication delay severity. The analytical solution, which is a function of the ratio between the communication delay and the local processing time, can be used to achieve the fastest convergence speed of the distributed dual coordinate ascent in time. Finally, we demonstrate the usability of our proposing algorithm in machine learning over communication networks, where local workers cannot directly reach to a central node.}


The rest of the paper is organized as follows. In Section \ref{sec:prob}, we introduce the regularized loss minimization problem with distributed data. Section \ref{sec:review} describes a review of existing works on the synchronous distributed dual coordinate ascent in a star network. In Section \ref{sec:propose}, we propose the generalized distributed dual coordinate ascent in tree-structured networks. Section \ref{sec:analysis} describes the convergence analysis of the generalized distributed dual coordinate ascent. In Section \ref{sec:optH}, we study the communication delay factor in the convergence speed of the distributed dual coordinate ascent. In Section \ref{sec:experiment}, we demonstrate the performance of the generalized distributed dual coordinate ascent and the optimal iteration numbers for the fast convergence speed.  The proposed algorithm and its convergence rate without a proof were introduced in our previous conference paper \cite{cho2019generalized}. \textcolor[rgb]{0,0,0}{In this journal paper, we provide the full proof of our theorem in Appendix \ref{appx:main_proof}, the analysis of network topology and communication effect on the algorithm in Sections \ref{sec:analysis} and \ref{sec:optH} respectively, and additional numerical experiments in Section  \ref{sec:experiment}.}


\textbf{Notations:} We denote the set of real numbers as $\R$. We use $[k]$ to denote the index set of the coordinates in the $k$-th coordinate block. For an index set $Q$, $\overline{Q}$ and $|Q|$ are used to represent the complement and the cardinality of $Q$ respectively. We use bold letters to represent vectors and matrices. If we use an index set as a subscript of a vector (resp. matrix), we refer to the partial vector (resp. partial matrix) over the index set (resp. with columns over the index set). The superscript $(t)$ is used to denote the $t$-th iteration. For example, $\balpha^{(t)}_{[k]}$ represents a partial vector $\balpha$ over the $k$-th block coordinate set at the $t$-th iteration. We reserve the superscript $\star$ to denote the optimal solution to an optimization problem.

\section{Problem formulation}
\label{sec:prob}
We consider the following regularized loss minimization problem \cite{yang2013trading,jaggi2014communication,shalev2013stochastic,huo2016distributed,hsieh2015PASSCoDe,ma2015adding}:
\par\noindent\small
\begin{align}\label{prob:primal}
    \underset{\bw \in \R^d}{\text{minimize}}\; P(\bw) \triangleq \frac{\lambda}{2} ||\bw||^2 + \frac{1}{m} \sum_{i=1}^{m} \ell_i(\bw^T \bx_i),
\end{align}
\normalsize
where $\bx_i \in \R^d$, $i=1,2,...,m$, are data points, $\ell_i(\cdot)$, $i=1,2,...,m$, are loss functions, and $\lambda$ is a tuning parameter for a regularization term. \textcolor[rgb]{0,0,0}{Note that due to the regularization term for $\bw$, which is a global variable, this minimization problem is not separable for each distributed node unlike the consensus problems introduced in  \cite{ying2019supervised,chang2015multi,shi2014linear,Mateos2010Distributed,Hong2017Stochastic}, where the regularization term is defined like $\sum_{k=1}^{K} r(\bw_k)$. Here $r(\cdot)$ is a regularization function and $K$ is the number of distributed nodes.} By considering different loss functions, \eqref{prob:primal} can be interpreted as various machine learning problems including regression and classification. For instance, for linear classification, by choosing the loss function $\ell_i(\cdot)$ to the hinge loss, i.e., $\ell_i(\bw^T\bx_i) = \max( 0, 1 - y_i (\bw^T \bx_i) )$,  \eqref{prob:primal} with labeled dataset $\{(\bx_i,y_i)\}_{i=1}^m$, where $y_i \in \R$ is label information, can be understood as the linear Support Vector Machine (SVM) classification problem. For regression, we can set $\ell_i(\bw^T \bx_i) = (\bw^T\bx_i - y_i)^2$ with some measurement data $y_i$, $i=1,2,...,m$. \textcolor[rgb]{0,0,0}{Throughout the paper, we assume that the data points $\bx_i$, $i=1,2,...,m$, are normalized in $\ell_2$ norm, i.e., $\| \bx_i \| \leq 1$, $i=1,2,...,m$, and the dataset $\{(\bx_i,y_i)\}_{i=1}^m$ is stored in a distributed manner over a  network having $K$ local workers. More specifically, the $k$-th local worker has training data $\{ (\bx_i, y_i) \}$, $i \in [k]$, where $[k]$ represents the index set for the training data of the $k$-th local worker. Hence, we have $|\cup_{k=1}^{K} [k]| = m$.}

From the primal problem \eqref{prob:primal}, we have the following dual problem by considering the conjugate function, i.e., $\ell_i(a)=\sup_{b} ab - \ell_i^{*}(b)$, where $a,b \in \R$:
\par\noindent\small
\begin{align}\label{prob:dual}
    \underset{\balpha \in \R^m}{\text{maximize}}\; D(\balpha) \triangleq -\frac{\lambda}{2} ||\bA\balpha||^2 - \frac{1}{m} \sum_{i=1}^{m} \ell^{*}_i(-\alpha_i),
\end{align}
\normalsize
where $\alpha_i$ is the $i$-th element of the dual vector $\balpha \in \R^m$, and the data matrix $\bA \in \R^{d \times m}$ whose $i$-th column is $\frac{1}{\lambda m} \bx_i$, i.e., $\bA_i = \frac{1}{\lambda m} \bx_i$,  is introduced for notation convenience. By defining $\bw(\balpha) \triangleq \bA \balpha$ shown in \cite{shalev2013stochastic, jaggi2014communication}, we have the duality gap as $P(\bw(\balpha)) - D(\balpha)$ for a useful and readily computable stopping criteria. It is noteworthy that from the duality principle \cite{boyd2004convex}, we have $P(\bw) \geq D(\balpha)$ for all $\bw$ and $\balpha$, and thus, $P(\bw(\balpha)) \geq D(\balpha)$ for all $\balpha$. If $\balpha = \balpha^{\star}$, which is the optimal solution to the dual problem \eqref{prob:dual}, and the loss function $\ell(\cdot)$ is convex, we have $P(\bw(\balpha^{\star})) = D(\balpha^{\star})$ from strong duality condition. Thus, $\bw(\balpha^{\star})$ becomes $\bw^{\star}$, which is the optimal solution to the primal problem \eqref{prob:primal}. 
\textcolor[rgb]{0,0,0}{Additionally, if the loss function $\ell_i(\cdot)$ is non-convex, the primal problem will become a non-convex problem. However, the dual problem is still expressed in a convex problem \cite{boyd2004convex}. Therefore, our algorithm to tackle the dual problem can provide an optimal solution to the dual problem. Unfortunately, in this case, there is no guarantee that  the optimal solution to the dual problem becomes an optimal solution to the primal problem.}

In the following sections, we consider a distributed dual coordinate ascent for the regularized loss minimization problem over distributed data. We firstly review the previous research on the distributed dual coordinate ascent in a star network.

\section{Review of the distributed dual coordinate ascent in a star network} \label{sec:review}
\begin{figure}[t]
    \centering
    \includegraphics[scale=0.3]{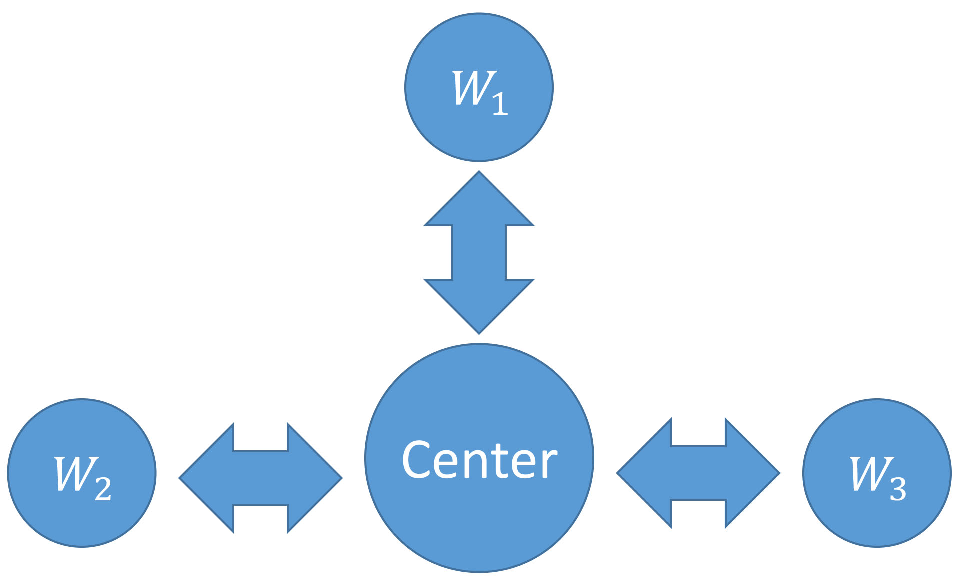}
    \caption{Illustration of a star network having  one central station and three local workers $W_1$, $W_2$ and $W_3$.}
    \label{fig:singleDistSystem}
\end{figure}

The distributed dual coordinate ascent for the regularized loss minimization problem over distributed data in a network has been studied in  \cite{yang2013trading,jaggi2014communication,huo2016distributed,ma2015adding}, where a star network topology for the network is considered as shown in Figure \ref{fig:singleDistSystem}. In particular, the authors in \cite{jaggi2014communication} introduced a distributed dual coordinate ascent framework, called the Communication-Efficient Distributed Dual Coordinate Ascent (CoCoA), and later proposed CoCoA+\cite{ma2015adding}, which is an enhanced version of CoCoA by adjusting the parameter value in the accumulation of intermediate results for faster convergence speed than CoCoA. Since we are interested in the distributed dual coordinate ascent for various structural network topologies and their influences to the performance of the distributed algorithm, we provide a high level review of CoCoA proposed in \cite{jaggi2014communication}.

Suppose a star network has $K$ local workers and each local worker has disjoint parts of dataset $\{ (\bx_i, y_i) \}_{i=1}^m$. With this problem setting, the authors in \cite{jaggi2014communication} introduced the distributed dual coordinate ascent for a star network. Due to the nature of the distributed algorithm, the algorithm updates the global variable in the outer iteration, and locally each worker has inner iterations. Particularly, at the $t$-th outer iteration of the algorithm, each worker solves a local dual problem for given dataset via LocalDualMethod($\cdot$), which represents any dual method to solve (\ref{prob:dual}), e.g. Stochastic Dual Coordinate Ascent (SDCA), simply denoted by LocalSDCA($\cdot$), through inner iterations. And then, each local worker sends the intermediate solution to the center node. The center node collects and accumulates all the results from the local workers, and then updates and shares the global solution $\bw^{(t)}$ at the $t$-th outer iteration back to the workers. Algorithm \ref{alg:CoCoA} describes the detail steps of the distributed coordinate ascent in a star network. The following theorem characterizes the convergence rate of the algorithm in \cite{jaggi2014communication}. 

\begin{algorithm}[t]
  \caption{Communication-efficient Distributed Dual Coordinate Ascent (CoCoA) \cite{jaggi2014communication}}
  \label{alg:CoCoA}
  \SetAlgoLined
{\small
   \KwIn{ $T \geq 1$}
   \KwOut{ $\bw$, $\balpha$ }
   \textbf{Data:} $\{(\bx_i, y_i)\}_{i=1}^{m}$ distributed over $K$ local workers \par
   \textbf{Initialization:} $\balpha_{[k]}^{(0)} \leftarrow \bm{0}$ for all local workers, and $\bw^{(0)} \leftarrow \bm{0}$ \par
   \For { $t=1$  \KwTo $T$ }
   {
       \For{ all local workers $k=1,2,...,K$ in parallel}
       {
        $(\bigtriangleup \balpha_{[k]}, \bigtriangleup \bw_{k})$ $\leftarrow$ LocalDualMethod($\balpha_{[k]}^{(t-1)}, \bw^{(t-1)}$) \par
        $\balpha_{[k]}^{(t)}$ $\leftarrow$ $\balpha_{[k]}^{(t-1)} + \frac{1}{K} \bigtriangleup \balpha_{[k]}$ \par
       }
       send $\bigtriangleup \bw_k$, $k=1,...,K$, to the central station \par
       $\bw^{(t)}$ $\leftarrow$ $\bw^{(t-1)} + \frac{1}{K} \sum_{k=1}^{K} \bigtriangleup \bw_k$ \par
       distribute $\bw^{(t)}$ to local workers \par
   }
}%
\end{algorithm}

\begin{theorem}[ {{\cite[Theorem 2]{jaggi2014communication}}} ]\label{thm:cocoa}
Suppose that Algorithm \ref{alg:CoCoA} is run for $T$ outer iterations of $K$ local computers with the procedure LocalSDCA($\cdot$) having local geometric improvement $\Theta$. Further, assume that the loss functions $\ell_i(\cdotp)$ are $\nicefrac{1}{\gamma}$-smooth. Then, the following geometric convergence rate holds for the global (dual) objective:
\par\noindent\small
\begin{align}\label{ieq:cocoa_conv}
    & \E[D(\balpha^{\star}) - D(\balpha^{(T)})]  \leq \big( 1-(1-\Theta) \frac{1}{K} \frac{\lambda m \gamma}{ \rho + \lambda m \gamma}\big)^T \big( D(\balpha^{\star})-D(\balpha^{(0)}) \big),
\end{align}
\normalsize
where $m$ is the size of the whole dataset and $\rho$ is any real number satisfying
\par\noindent\small
\begin{align*}
    \rho \geq \rho_{min} \triangleq \underset{\balpha \in \mathbb{R}^m}{\text{maximize}}\;\; \lambda^2 m^2 \frac{\sum_{k=1}^{K} ||\bA_{[k]}\balpha_{[k]} ||^2 - ||\bA\balpha||^2  }{ ||\balpha||^2} \geq 0.
\end{align*}
\normalsize
\end{theorem}

With LocalSDCA($\cdot$), which uses the SDCA to solve the dual problem for given dataset at each worker, the local geometric improvement $\Theta$ can be set to
\par\noindent\small
\begin{align}\label{eq:Theta}
    \Theta = (1 - s/\tilde{m})^H,
\end{align}
\normalsize
where $\tilde{m} \triangleq \max_{k=1,...,K} m_k$ is the size of the largest block of coordinates among $K$ local workers, $H$ is the number of local (or inner) iterations in LocalSDCA($\cdot$), and $s \in [0,1]$ is a step size of the gradient ascent which determines how far the next solution will be taken from the current solution at each iteration. Additionally, by choosing different parameter values instead of $\frac{1}{K}$ in the summation of $\bigtriangleup \bw_k$'s in Algorithm \ref{alg:CoCoA}, the authors in \cite{ma2015adding} proposed CoCoA+, which has the same framework as CoCoA introduced in Algorithm \ref{alg:CoCoA}, for faster convergence speed than CoCoA. 

CoCoA has been shown to work well for distributed machine learning problems with distributed data in a star network, which is a simple network model. However, the topology of a network may not necessarily be a star network. In the next section, we study the distributed dual coordinate ascent in a general network, which is a tree-structured network model.

\section{Generalized distributed dual coordinate ascent in tree networks}
\label{sec:propose}
One may think of a connected communication network, e.g., a spanning tree network, as a virtual star network by considering the long relays of links from a central node to each leaf node as a direct virtual-link through intermediate nodes. \textcolor[rgb]{0,0,0}{However, since communication delays normally exist in a network and the communication is a big burden of distributed algorithms,} the distributed algorithms in the virtual star network can easily suffer from the long delays in communication by significantly slowing down the convergence of the distributed algorithms. Therefore, in a connected communication network, it is efficient to perform distributed optimization among local workers close to each other, and then, communicate the intermediate results to a central or sub-central nodes. Based on this idea, we investigate how to design the distributed dual coordinate ascent over a general tree-structured network, and provide its convergence analysis. Since every connected network has a spanning tree, we choose to investigate the distributed algorithm over a tree-structured network, which is also a generalization of a star network.

\begin{figure}[t]
    \centering
    \includegraphics[scale=0.35]{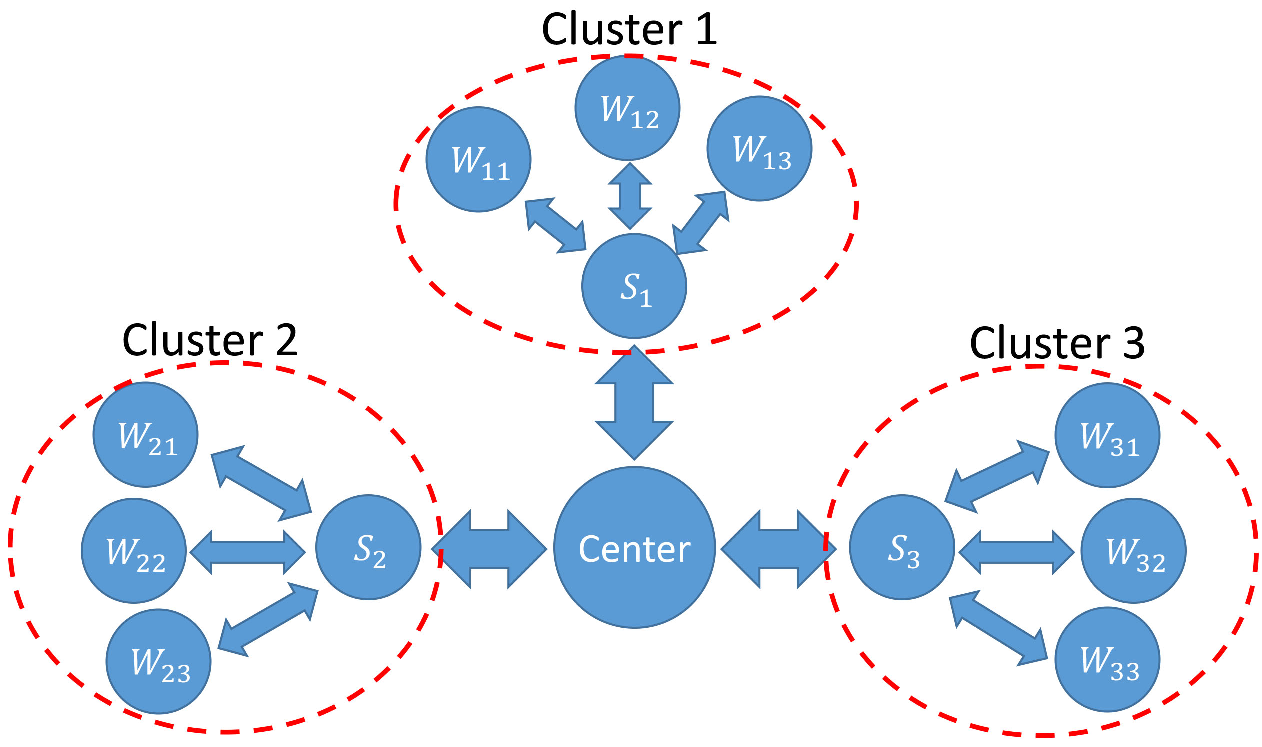}
    \caption{Illustration of a tree-structured network, which has two layers. In the network, a central station (root node) has three direct child nodes $S_1$, $S_2$ and $S_3$. Each node $S_i$ has three direct child nodes $W_{ij}$, $j=1,2,3$. }
    \label{fig:generalDistSystem}
\end{figure}

\textcolor[rgb]{0,0,0}{In Figure \ref{fig:generalDistSystem}, we show a two-layer tree network as an example of a general tree-structured network, where the number of layers represents the depth of the tree network. The root node of the tree network represents the central station of the network. Each tree node may have several direct child nodes. For example, the root node has three direct child nodes $S_1$, $S_2$, and $S_3$ in Figure \ref{fig:generalDistSystem}. A node not having any child node is called as a leaf node. Without loss of generality, we assume that only leaf nodes have the distributed data, which are disjoint segmented blocks of the data matrix $\bA$ in column-wise. Note that $\bA_i = \frac{1}{\lambda m} \bx_i$, where $\bA_i$ is the $i$-th column of $\bA$ and $\bx_i \in \R^{d}$ is the $i$-th data point. If a non-leaf node $Q$ has data, we can always create a virtual leaf node $L$ attached to $Q$, and ``stores'' the data in $L$. Thus, without loss of generality, we can assume that the dataset $\{ (\bx_i , y_i) \}_{i=1}^{m}$ are distributed only to leaf nodes. }

\begin{figure}[t]
    \centering
    \includegraphics[scale=0.45]{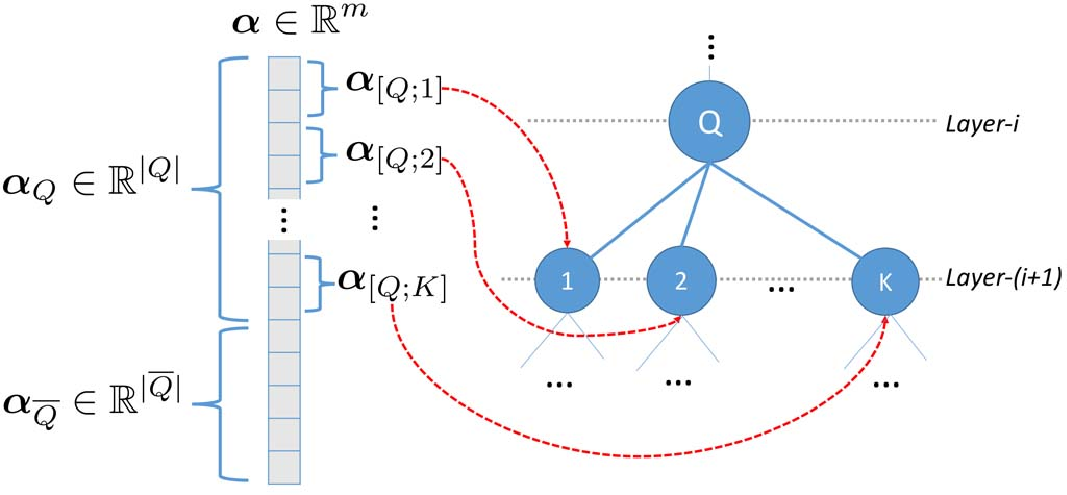}
    \caption{Illustration of a subtree including a node $Q$ on the $i$-th layer and its direct and indirect child nodes.}
    \label{fig:subtreeQ}
\end{figure}

\textcolor[rgb]{0,0,0}{For a tree node $Q$, we can consider a subtree including the tree node $Q$ and its indirect and direct child nodes up to leaf nodes, simply called the subtree $Q$. Figure \ref{fig:subtreeQ} illustrates the subtree $Q$. We also denote the set of indices of all data points stored in the subtree $Q$ as $Q$, and the set of indices of data points in the $k$-th direct child node of $Q$ as $[Q;k]$. Therefore, $[Q;k] \subset Q$. Then, $\balpha_{[Q;k]}$ represents the partial vector of $\balpha \in \R^m$ corresponding to the data points in the subtree with the $k$-th direct child node of $Q$. Since each node is used for an index set, we denote the number of data points stored in the subtree $Q$, i.e., the cardinality of $Q$, as $|Q|$. In a tree network, we also assume that a node can only communicate with its direct child nodes or its direct parent node. }

\textcolor[rgb]{0,0,0}{We then introduce the generalized distributed dual coordinate ascent, which we call TreeDualMethod, to solve the dual problem (\ref{prob:dual}) with distributed data stored over a general tree-structured network. For simplicity, we consider the tree network in Figure \ref{fig:generalDistSystem}, where the number of layers, $p$, is 2. In a leaf node $W_{ij}$, TreeDualMethod in Procedure \ref{alg:leaf} is run with a local dataset for $T_p$ iterations, and then, the intermediate value $\bigtriangleup \bw$ is shared with its direct parent node, i.e., the sub-central node $S_i$. In the sub-central node $S_i$, a global variable $\bw$ for the $i$-th cluster is updated, and distributed to local workers $W_{ij}$'s. After running this process for $T_i$ times independently in clusters, the variables $\bigtriangleup \bw$'s from clusters are shared with the central node. The central node updates and shares the global variable $\bw$ for whole distributed nodes. And the algorithm repeats this process until some stopping criteria holds. Algorithm \ref{alg:root}, Algorithm  \ref{alg:general} and Procedure \ref{alg:leaf}  describe the computational steps of TreeDualMethod for the root node, a general tree node (not root or leaf),  and a leaf node respectively.}

\textcolor[rgb]{0,0,0}{It is noteworthy that like the distributed algorithm in a star network case, in the distributed networks, the output $\bigtriangleup \bw_Q$ in Procedure \ref{alg:leaf} and Algorithm \ref{alg:general} or the output $\bw$ in Algorithm \ref{alg:root} are transmitted between nodes, while the outputs $\balpha$ and $\bigtriangleup \balpha_Q$ are not transmitted through communication networks. Each node generates $\balpha$ or $\Delta \balpha_Q$ as an output of each node, but those outputs are used in each node at the next iteration without transmission to other nodes. Therefore, even though we have a large dataset, the communication cost is not affected by the size of the dataset. Also, when the dimension of $\balpha \in \R^m$ is large, i.e., large amount of data, and the dimension of  $\bw$ is much smaller than $m$, which is normally the case in big data, our distributed algorithm will have less communication burden. }

\begin{algorithm}[t]
  \caption{TreeDualMethod: Distributed Dual Coordinate Ascent for the Root Node $Q$ on the layer-$0$}
  \label{alg:root}
  \SetAlgoLined
{\small
   \KwIn{ $T_0 \geq 1$}
   \textbf{Initialization}: $\balpha_{[Q; k]}^{(0)} \leftarrow 0$ for all direct child nodes $k$ of node $Q$,  $\bw^{(0)} \leftarrow 0$   \par
   \For { $t=1$  \KwTo $T_0$ }
   {
       \For{ all direct child nodes $k=1,2,...,K_0$ in parallel}
       {
            $(\bigtriangleup \balpha_{[Q;k]}, \bigtriangleup \bw_{k})$ $\leftarrow$ TreeDualMethod($\balpha_{[Q;k]}^{(t-1)}, \bw^{(t-1)}$) \par
            $\balpha_{[Q;k]}^{(t)}$ $\leftarrow$ $\balpha_{[Q;k]}^{(t-1)} + \frac{1}{K_0} \bigtriangleup\balpha_{[Q;k]}$ \par
       }%
       $\bw^{(t)}$ $\leftarrow$ $\bw^{(t-1)} + \frac{1}{K_0} \sum_{k=1}^{K_0} \bigtriangleup \bw_k$ \par
   }%
  \KwOut{ $\balpha^{(T_0)}$,  and $\bw^{(T_0)}$ }
}%
\end{algorithm}

\begin{algorithm}[t]
  \caption{TreeDualMethod: Distributed Dual Coordinate Ascent for a  General Tree Node $Q$ on the layer-$i$, $i=1,2,...,p-1$}
  \label{alg:general}
  \SetAlgoLined
{\small
   \KwIn{ $T_i \geq 1$, $\balpha_{Q}$,  $\bw$}
   \textbf{Initialization}: $\balpha_{[Q;k]}^{(0)} \leftarrow \balpha_{[Q;k]}$ for all direct child nodes $k$ of node $Q$ ,  $\bw^{(0)} \leftarrow \bw$   \par
   \For { $t=1$  \KwTo $T_i$ }
   {\For{ all direct child nodes $k=1,2,...,K_i$ of $Q$ in parallel}
       {
        $(\bigtriangleup \balpha_{[Q;k]}, \bigtriangleup\bw_{k})$ $\leftarrow$ TreeDualMethod($\balpha_{[Q;k]}^{(t-1)}, \bw^{(t-1)}$) \par
        $\balpha_{[Q;k]}^{(t)}$ $\leftarrow$ $\balpha_{[Q;k]}^{(t-1)} + \frac{1}{K_i} \bigtriangleup\balpha_{[Q;k]}$ \par
       }%
       $\bw^{(t)}$ $\leftarrow$ $\bw^{(t-1)} + \frac{1}{K_i} \sum_{k=1}^{K_i} \bigtriangleup \bw_k$ \par
} %
  \KwOut{$\bigtriangleup\balpha_{Q} \triangleq \balpha_{{Q}}^{(T_i)}-\balpha_{{Q}}^{(0)}$, and $\bigtriangleup \bw_{Q} \triangleq \bA_{Q} \bigtriangleup \balpha_{Q}$  }
}%
\end{algorithm}


\begin{procedure}[t]
\caption{P(). TreeDualMethod: Distributed Dual Coordinate Ascent for a Leaf Node $Q$ on the layer-$p$}
  \label{alg:leaf}
  \SetAlgoLined
{\small
   \KwIn{ $T_p \geq 1$, $\balpha_{Q} \in \R^{|Q|}$, and $\bw \in \R^{d}$ consistent with other coordinate blocks of $\balpha$ s.t. $\bw=\bA\balpha$ }
   \textbf{Data:} ${\{(\bx_i, y_i)\}}_{i \in Q}$  \par
   \textbf{Initialization:} $\bigtriangleup \balpha_{Q} \leftarrow 0 \in \R^{|Q|}$, and $\bw^{(0)} \leftarrow \bw$ \par
   \For { $h=1$  \KwTo $T_p$ }
   {    choose $i \in Q$ uniformly at random \par
        find $\bigtriangleup \alpha$ maximizing $-\frac{\lambda m}{2} ||\bw^{(h-1)} + \frac{1}{\lambda m} \bigtriangleup\alpha \bx_i||^2 - \ell_i^{*}(-(\alpha_i^{(h-1)} + \bigtriangleup\alpha))$ \par
        $\alpha_{i}^{(h)}$ $\leftarrow$ $\alpha_{i}^{(h-1)} + \bigtriangleup \alpha$ \par
        $ (\bigtriangleup \balpha_{Q})_i$ $\leftarrow$ $(\bigtriangleup\balpha_{Q})_i + \bigtriangleup\alpha$ \par
	$\bw^{(h)}$ $\leftarrow$ $\bw^{(h-1)} + \frac{1}{\lambda m} \bigtriangleup\alpha \bx_i$ \par
   }
      \KwOut{ $\bigtriangleup\balpha_{Q}$ and $\bigtriangleup\bw_{Q} \triangleq \bA_{Q} \bigtriangleup\balpha_{Q}$ }
}%
\end{procedure}

\section{Convergence analysis of TreeDualMethod over a tree network}
\label{sec:analysis}

We analyze the convergence rate of the distributed dual coordinate ascent in a general tree-structured network model in this section. In a nutshell, we will show a recursive relation between the convergence rate of the algorithm at a tree node $Q$ and that at the node $Q$'s direct child nodes. Hence, the overall convergence rate of the distributed dual coordinate ascent in a general tree-structured network can be understood in a recursive manner, where the number of recursions is dependent on the number of layers of the tree network.

For clear description, let us consider a general tree network model having $p$ layers from the root node to leaf nodes, where the root node is on the layer-$0$ and the leaf nodes are on the layer-$p$. Suppose a node $Q$ on the $i$-th layer has $K$ direct child nodes on the $(i+1)$-th layer shown in Figure \ref{fig:subtreeQ}. We use $\balpha_{[Q; k]}$ to denote the partial dual variable vector corresponding to its $k$-th direct child node, where $1\leq k \leq K$. Then, let us define the local suboptimality gap for the $k$-th direct child node of $Q$ as
\par\noindent\small
\begin{align}\label{def:epsilon_Qk}
     & \epsilon_{Q,k}(\balpha) \triangleq \underset{\hat{\balpha}_{[Q;k]}}{\text{maximize}} \;D(\balpha_{[Q;1]},...,\hat{\balpha}_{[Q;k]},...,\balpha_{[Q;K]},\balpha_{\overline{Q}} ) - D(\balpha_{[Q;1]},...,\balpha_{[Q;k]},...,\balpha_{[Q;K]}, \balpha_{\overline{Q}}).
\end{align}
\normalsize
Remark that the local suboptimality gap for the $k$-th child node is defined with fixing $\balpha_{\overline{Q}}$ and $\balpha_{[Q;i]}$'s, where $i\neq k$, and only updating $\balpha_{[Q;k]}$.  Thus, the local suboptimality gap for the $k$-th direct child node of $Q$ represents the maximum objective value gap that the $k$-th direct child node of $Q$ can achieve from the current $\balpha^{(t)}$ value with fixing other $\alpha_i$, $i \notin [Q;k]$, variables. Then, we introduce the following assumption about the local geometric improvement of TreeDualMethod at the $k$-th direct child node of $Q$.

\begin{assumption} [Geometric improvement of TreeDualMethod at a direct child node]
\label{asp:local_assumption}
For a tree node $Q$ \textcolor[rgb]{0,0,0}{on the $i$-th layer,} we assume that there exists $\Theta_{i+1} \in [0,1)$ such that for any given $\balpha$, TreeDualMethod at the $k$-th direct child node of $Q$ returns an update $\bigtriangleup \balpha_{[Q;k]}$ satisfying 
\par\noindent\small
\begin{align} \label{eq:Theta_i1}
    & \E[ \epsilon_{Q,k} (\balpha_{[Q;1]},...,\balpha_{[Q;k-1]}, \balpha_{[Q;k]}+\bigtriangleup \balpha_{[Q;k]},...,\balpha_{[Q;K]}, \balpha_{\overline{Q}})]  \leq \Theta_{i+1} \cdot \epsilon_{Q,k}(\balpha).
\end{align}
\normalsize
\end{assumption}
\noindent\textcolor[rgb]{0,0,0}{Note that Assumption \ref{asp:local_assumption} here is introduced for an arbitrary tree node in a general tree network and used as a starting assumption in mathematical induction for recursive convergence analysis,} while Assumption 1 of \cite{jaggi2014communication} is introduced for an abstract function in the distributed algorithm framework.

\begin{figure*}[t]
    \centering
    \includegraphics[scale=0.65]{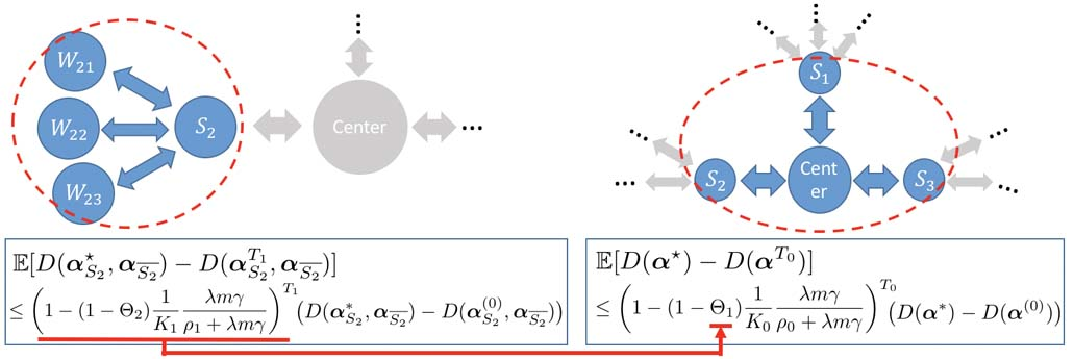}
    \caption{Illustration of the structure of the tree network factor in convergence analysis.}
    \label{fig:recursion}
\end{figure*}

For a leaf node, we use LocalSDCA for TreeDualMethod described in Procedure \ref{alg:leaf} as in \cite{jaggi2014communication}, and provide the following proposition about the convergence bound for a leaf node $B$ even with the input $\bw$ also determined by $\balpha_{\overline{Q}}$ and $\balpha_{Q\setminus B}$ in Procedure \ref{alg:leaf}.
\begin{proposition}[{\cite[Proposition 1]{jaggi2014communication}} ]
Let us consider a tree node $Q$ whose direct child node $B$ is a leaf node. Assume that loss functions $\ell_i(\cdot)$ are $\nicefrac{1}{\gamma}$-smooth. Then for the leaf node $B$, Assumption \ref{asp:local_assumption} holds with
\par\noindent\small
    \begin{align}\label{eq:theta_p}
        \Theta_p=\big( 1- \frac{\lambda m \gamma}{1 + \lambda m \gamma} \frac{1}{{m_{B}}}\big)^{T_p}.
    \end{align}
\normalsize
    where $m_{B}$ is the size of data stored at node $B$, $T_p$ is the number of iterations in Procedure \ref{alg:leaf}. 
    \label{leafconver}
\end{proposition}
\noindent \textcolor[rgb]{0,0,0}{Basically, the geometric improvement condition holds true with LocalSDCA if the $k$-th direct child node of $Q$ is a leaf child node with $\Theta$ introduced in \eqref{eq:Theta}, where $s$ in \eqref{eq:Theta} is $\frac{\lambda m \gamma}{1 + \lambda m \gamma}$ in \eqref{eq:theta_p}.}

Additionally, Theorem \ref{recur_conver}, which is our main result, shows that if the geometric improvement condition holds true for direct child nodes of $Q$, then the geometric improvement condition also holds true for $Q$; thus it leads to a recursive calculation of the convergence rate for the whole tree network.

\begin{theorem}
\label{recur_conver}
Let us consider a tree node $Q$ on the $i$-th layer which has $K_i$ direct child nodes satisfying the local geometric improvement requirement introduced in Assumption \ref{asp:local_assumption}, with parameters $\Theta^{1}_{i+1}$, $\Theta_{i+1}^2$, ..., and $\Theta^{K_i}_{i+1}$. We assume that Algorithm \ref{alg:general} (or Algorithm \ref{alg:root}) has an input $\bw$ and is run for $T_i$ iterations. We further assume that loss functions $\ell_i(\cdot)$'s are $\nicefrac{1}{\gamma}$-smooth.

Then, for any input $\bw$ to Algorithm \ref{alg:general} (or Algorithm \ref{alg:root}), the following geometric convergence rate holds for $Q$:
\par\noindent\small
\begin{align}
\label{ieq:convergence}
    & \E [D(\balpha_Q^{*}, \balpha_{\overline{Q}}) - D(\balpha_Q^{(T_i)}, \balpha_{\overline{Q}}) ]  \leq \big( 1-(1-\Theta_{i+1}) \frac{1}{K_i} \frac{\lambda m \gamma}{ \rho_i + \lambda m \gamma}\big)^{T_i} \big(D(\bm{\alpha}_Q^{*}, \bm{\alpha}_{\overline{Q}})-D(\balpha_Q^{(0)},\balpha_{\overline{Q}}) \big),\nonumber 
\end{align}
\normalsize
where $\Theta_{i+1} = \max_{k} \Theta^{k}_{i+1}$, and $\rho_i$ is any real number satisfying
\par\noindent\small
\begin{align*}
    \rho_i \geq \rho_{min} \triangleq \underset{\balpha_Q \in \mathbb{R}^{|Q|}}{\text{maximize}}\;\; \lambda^2 m^2 \frac{\sum_{k=1}^{K_i} ||\bA_{[Q;k]}\balpha_{[Q;k]} ||^2 - ||\bA_{Q}\balpha_{Q}||^2  }{ ||\balpha_{Q}||^2} \geq 0.
\end{align*}
\normalsize
\end{theorem}
\noindent \textcolor[rgb]{0,0,0}{Note that the parameter $\rho_i$ is related to the overlapping level among the datasets in the subtree $Q$. When we have more overlap among local datasets in the subtree, the parameter $\rho_i$ can become larger, which will lead to slower convergence rate.} 

Proposition \ref{leafconver} is for the local geometric improvement of TreeDualMethod at a leaf node. \textcolor[rgb]{0,0,0}{Namely, Assumption 1 holds for leaf nodes.} Theorem \ref{recur_conver} is for the local geometric improvement of TreeDualMethod at any non-leaf tree node. Note that $( 1-(1-\Theta_{i+1}) \frac{1}{K_i} \frac{\lambda m \gamma}{ \rho_i + \lambda m \gamma} )^{T_i}$ in \eqref{ieq:convergence} becomes $\Theta_i$ for a tree node $Q$ on the $i$-th layer, and then, (\ref{ieq:convergence}) is interpreted as the local geometric improvement of TreeDualMethod at the direct child node by the direct parent node of $Q$, which is a node on the $(i-1)$-th layer. \textcolor[rgb]{0,0,0}{Basically, for the convergence rate of the generalized dual coordinate ascent over the whole tree network, we use the mathematical induction, where Proposition \ref{leafconver} is the base case,  Assumption 1 is the starting assumption of the mathematical induction, and Theorem \ref{recur_conver} completes the induction for the recursive convergence analysis. Therefore, by combining Theorem \ref{recur_conver} with Proposition 1, we can recursively obtain the convergence rate of the generalized distributed dual coordinate ascent algorithm for the whole tree network with the fact that Assumption 1 holds true for every node in a tree network.} \textcolor[rgb]{0,0,0}{Figure \ref{fig:recursion} illustrates the structure of the tree network factor in convergence rate, shown through $\Theta_{1}$ and  $\Theta_{2}$.}  

\textcolor[rgb]{0,0,0}{We remark that Theorem \ref{recur_conver} is different from Theorem 2 of \cite{jaggi2014communication} in three aspects. Firstly, Theorem \ref{recur_conver} is applicable to any tree node in a general tree network, beyond a star network discussed in \cite{jaggi2014communication}. Secondly, even when the input $\bw$ of Algorithm \ref{alg:general} is determined by not only $\balpha_{Q}$ but also $\balpha_{\overline{Q}}$, Theorem \ref{recur_conver} holds. Note that $\bw = \bA(\balpha_Q, \balpha_{\overline{Q}} ) = \bA_Q\balpha_Q + \bA_{\overline{Q}} \balpha_{\overline{Q}}$. Unlike our Theorem, in Theorem 2 of \cite{jaggi2014communication}, due to the star network topology, a local worker has $\bw$ as an input from the root node which is updated with intermediate results obtained from all the local workers. Hence, $\balpha_{\overline{Q}}$ is not considered in Theorem 2 of \cite{jaggi2014communication} and its proof. Our proof of Theorem \ref{recur_conver} addresses this challenge that the input $\bw$ is also affected by $\balpha_{\overline{Q}}$.   Therefore, we have to deal with both updating coordinates $\balpha_{Q} \in \R^{|Q|}$ and un-updating coordinates $\balpha_{\overline{Q}} \in \R^{| \overline{Q}|}$, where $|Q| + |\overline{Q}|  = m$, while in the proof of Theorem 2 of  \cite{jaggi2014communication}, all the coordinates are updating coordinates, i.e., $\balpha \in \R^{m}$. For the readability, we place the proof of Theorem \ref{recur_conver} in Appendix \ref{appx:main_proof}. Finally, unlike \cite{jaggi2014communication}, we do not consider the different local-dual problem introduced in Eqn. (8) of \cite{jaggi2014communication} for local workers, but deal with the original dual problem introduced in \eqref{prob:dual} with fixed $\overline{\bw} \triangleq \bA_{\overline{Q}}\balpha_{\overline{Q}}$ for a general tree node $Q$. Therefore, our theorem works for any tree node in a general tree network rather than just for one central node, which allows the recursive convergence analysis of the distributed dual coordinate ascent in a general tree network.}

By denoting the convergence bound in \eqref{ieq:convergence} as $\Theta_i$, i.e.,
\par\noindent\small
\begin{align}
	\Theta_{i} = \bigg( 1-(1-\Theta_{i+1}) \frac{C_i}{K_i} \bigg)^{T_i},
\end{align}
\normalsize
where $C_i$ represents $\frac{\lambda m \gamma}{ \rho_i + \lambda m \gamma}$, $T_i$ is the outer iteration in a tree node on the $i$-th layer, and $K_i$ is the number of direct child nodes attached to a tree node on the $i$-th layer, we can express the convergence bound on the whole tree-network, i.e., $\Theta_0$, in terms of the number of layers $p$, and the number of nodes $K_i$'s and $C_i$'s, as follows:
\par\noindent\small
\begin{align}\label{eq:Theta_0}
\Theta_0 & = \bigg( 1-\bigg(1-\Theta_{1} \bigg) \frac{C_0}{K_0} \bigg)^{T_0}  = \bigg( 1- \bigg(1- \bigg( 1- \cdots \bigg(1-\Theta_{p} \bigg) \frac{C_{p-1}}{K_{p-1}} \bigg)^{T_{p-1}} \bigg)\cdots \frac{C_1}{K_1} \bigg)^{T_1} \bigg) \frac{C_0}{K_0} \bigg)^{T_0}, 
\end{align}
\normalsize
where $\Theta_p$ is introduced in \eqref{eq:theta_p}. For simplicity, we assume that all tree nodes on the $i$-th layer have the same number of direct child nodes $K_i$.  

If $| T_i \cdot \frac{C_i \Theta_{i+1}}{K_i - C_i} | \ll 1$, for $i=0,1,...,p-1$, by applying the binomial approximation, we can have
\par\noindent\small 
\begin{align*}
	 \Theta_i &  = \bigg( \frac{K_i - C_i}{K_i} \bigg)^{T_i} \bigg( 1 + \frac{C_i}{K_i - C_i} \Theta_{i+1}   \bigg)^{T_i}  \approx \bigg( \frac{K_i - C_i}{K_i} \bigg)^{T_i} \bigg( 1+  \frac{C_i}{K_i - C_i} \Theta_{i+1} \cdot T_i. \bigg)
\end{align*}
\normalsize
and approximate \eqref{eq:Theta_0} as follows:
\par\noindent\small
\begin{align}\label{eq:Theta_0_approx}
	\Theta_0 \approx  & \bigg(\frac{K_0 - C_0}{K_0}\bigg)^{T_0} + \sum_{r=1}^{p-1} \prod_{i=0}^r \bigg( \frac{K_i - C_i}{K_i}\bigg)^{T_i} \cdot \prod_{j=0}^{r-1} \frac{C_j T_j}{ K_j - C_j} + \prod_{i=0}^{p-1} \bigg( \frac{K_i - C_i }{ K_i } \bigg)^{T_i} \bigg( \frac{C_i T_i}{K_i - C_i} \bigg) \Theta_p.
\end{align}
\normalsize
In Section \ref{sec:experiment}, we will investigate the gap between \eqref{eq:Theta_0} and \eqref{eq:Theta_0_approx} through numerical experiments as well as analyze the network topology's effect including the number of workers $K_i$ and the number of layers $p$ on the convergence bounds over the whole tree network introduced in \eqref{eq:Theta_0} and \eqref{eq:Theta_0_approx}.

We have discussed how the network topology can affect the convergence rate of the distributed dual coordinate ascent, which is expressed in terms of the number of layers, the number of nodes, and the number of iterations. However, for distributed algorithms, communications in a network can be a bottleneck of the convergence of the distributed algorithms. Therefore, it is quite natural to consider communication delay, which is normally expressed in time, in order to predict or estimate the convergence speed of the distributed algorithms. In the next section, we will study how communication delay, which is one of major network constraints, impacts the convergence of distributed dual coordinate ascent algorithms. By taking communication delays into account, we will optimize the number of local iterations $T_p$ in Procedure \ref{alg:leaf} and $T_i$ in Algorithm \ref{alg:general} for maximum convergence speed.

\section{Impacts of communication delay on the convergence rate of distributed dual coordinate ascent}
\label{sec:optH}
Earlier works \cite{yang2013trading,jaggi2014communication,ma2015adding} bounded  the convergence of distributed dual coordinate ascent algorithms with respect to the number of inner and outer iterations. However, in distributed algorithms, there may be significant communication delays in a distributed network. Thus, the convergence speed of distributed algorithms depends on not only  how many iterations of these algorithms have been run, but also the communication delays in performing these iterations. Intuitively, if the communication delay is close to zero, local workers may be better to perform a small number of local iterations, and communicate with the central station at a higher frequency; on the other hand, if the communication delay is large, namely, there is a large communication cost, then local workers may want to perform more local iterations before communicating with the central station in order to speed up convergence.  Therefore, our goal here is to investigate the convergence speed of distributed dual coordinate ascent with respect to total execution time including computational time and communication delays, and to optimize the number of local iterations by considering communication delays to achieve the maximum convergence speed of the distributed dual coordinate ascent. \textcolor[rgb]{0,0,0}{The research \cite{tsianos2012communication,Nedic2018Network,doan2017impact,Tsianos2013Role} studied the impact of the communication delays on the convergence rate of algorithms in various distributed optimization problems including distributed consensus problems. However, for the regularized loss minimization problem that we deal with in this paper, to the best of our knowledge, our paper is the first one to analytically study the communication delay's impact on the convergence rate, and finds the optimal number of local iterations depending on the communication delay severity.}

\begin{figure}[t]
    \centering
    \includegraphics[scale=0.40]{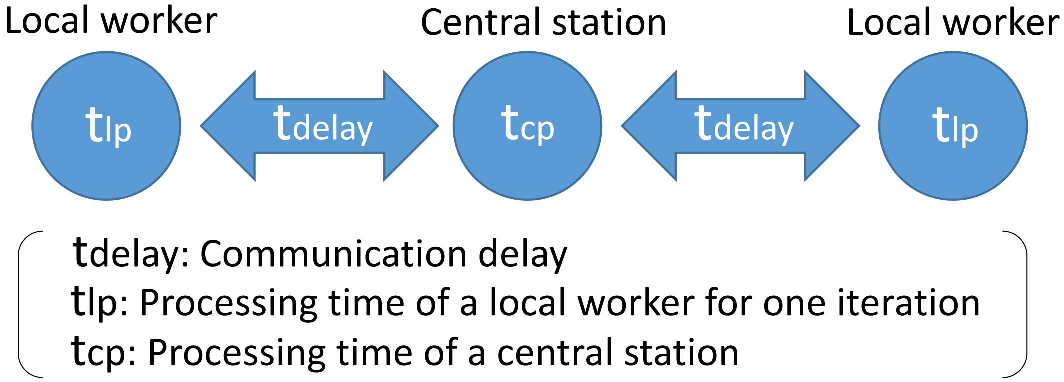}
    \caption{Definition of delay and computational time.}
    \label{fig:def_delay}
\end{figure}

For simplicity, let us first consider a star network as shown in Figure \ref{fig:singleDistSystem} and the corresponding Algorithm \ref{alg:CoCoA}. Since the communication delay is normally given in time, we need to consider both time and the number of iterations in the convergence analysis in order to obtain the optimal number of iterations in practical applications having communication delay and computational time. We denote the round-trip communication delay between a local worker and the central station as $t_{{delay}}$. We use $t_{lp}$  to denote the computational time for one local iteration at a worker, and use $t_{{cp}}$ to denote the computational time for parameter update at the central station. Figure \ref{fig:def_delay} illustrates the communication delay, and the processing time of each local and central station.

Suppose that each local worker performs $T_p$ local iterations before communicating with the central station, and there are $T_0$ outer iterations in total. Then, the total experienced time is 
\par\noindent\small
\begin{align}
    t_{total} = ( t_{lp} T_p + t_{delay} + t_{cp}) \cdot T_0.
\end{align}
\normalsize
Hence, the number of outer iterations $T_0$ is given by
\par\noindent\small
\begin{align}
\label{eq:iter_time}
    T_0=\frac{t_{total}}{(t_{lp} T_p + t_{delay} + t_{cp})}.
\end{align}
\normalsize

From (\ref{ieq:convergence}), for $T_0$ outer iterations, the expected gap between the optimal objective value and the current objective value with Algorithm \ref{alg:CoCoA} is expressed as
\par\noindent\small
\begin{align}
\label{eq:conv_iter}
    \bigg( 1-\big(1-[1 - \delta ]^{T_p} \big) \frac{C}{K} \bigg)^{T_0},
\end{align}
\normalsize
where $\delta = \frac{s}{\tilde{m}}$, $C=\nicefrac{\lambda m \gamma}{(\rho + \lambda m \gamma)}$, and $K$ is the number of local workers. In order to minimize the gap in objective value  (\ref{eq:conv_iter})  for a given total time $t_{{total}}$, we introduce the following optimization problem over the number of local iterations $T_p$ by plugging (\ref{eq:iter_time}) into (\ref{eq:conv_iter}):
\par\noindent\small
\begin{align}
\label{prob:primal_iter}
    \underset{T_p \geq 0}{\text{minimize}}\;  F(T_p) \triangleq \bigg( 1-\big(1-\big[1 - \delta \big]^{T_p} \big) \frac{C}{K} \bigg)^{ \frac{t_{total}}{t_{lp} T_p + t_{delay} + t_{cp}} }.
\end{align}
\normalsize

In order to figure out the optimal number of local iterations, let us find the critical point of the objective function $F(T_p)$. By applying logarithm to $F(T_p)$, we have 
\par\noindent\small
\begin{align}\label{eq:log_opt_iter}
\hspace{-0.2em} \ln F(T_p)  = \underbrace{ \frac{ t_{total}/ t_{lp} }{T_p + ( t_{delay} + t_{cp})/ t_{lp}  } }_{(A)} \underbrace{ \ln \bigg(\frac{K-C}{K}+\frac{C}{K}\big[1 - \delta \big]^{T_p} \bigg) }_{(B)}.
\end{align}
\normalsize
\eqref{eq:log_opt_iter} can be interpreted as the multiplication of two parts: the fraction part $(A)$ and the logarithm part $(B)$. Note that the fraction part $(A)$ is a decreasing function over $T_p$. And for the logarithm part $(B)$, as $T_p$ increases, $(B)$ goes to $\ln ((K-C)/K)$, which is less than zero, due to the condition $ 0 \leq 1-\delta < 1$. At $T_p=0$, $\ln F(T_p)$ is 0 due to $(B) = 0$. As $T_p$ goes to infinity, $\ln F(T_p)$ will go to 0 due to $(A) = 0$. Therefore, we can expect at least a critical point at some $T_p$. In order to figure out the critical point of \eqref{eq:log_opt_iter}, which is the same critical point of $F(T_p)$, we calculate the first order condition as follows:
\par\noindent\small
\begin{align}\label{eq:first_order}
\frac{d \ln F(T_p)}{d T_p} = & \frac{ (\frac{K-C}{K})  ( \frac{t_{total}}{ t_{lp} } ) (1-\delta)^{T_p} \ln (1- \delta)}{ \big(\frac{K-C}{K}+\frac{C}{K}\big[1 - \delta \big]^{T_p}  \big) \big( T_p + \frac{ t_{delay} + t_{cp}}{ t_{lp}} \big) }  - \frac{  ( \frac{ t_{total}} {t_{lp} }  ) \ln \big(\frac{K-C}{K}+\frac{C}{K}\big[1 - \delta \big]^{T_p}  \big) }{ \big(T_p + \frac{ t_{delay} + t_{cp}}{ t_{lp} }\big)^2  }  = 0.
\end{align}
\normalsize
By simplifying \eqref{eq:first_order} and denoting $\frac{t_{delay} + t_{cp}}{t_{lp}}$ to $r$,  we have the first order condition over $T_p$ as
\par\noindent\small
\begin{align}\label{eq:first_order2}
	&\underbrace{ \frac{K-C}{K}  \big( T_p + r \big) \big[1-\delta \big]^{T_p} \ln (1-\delta)}_{(C)}  
	- \underbrace{ \bigg(\frac{K-C}{K}+\frac{C}{K} \big[1-\delta\big]^{T_p} \bigg) \ln \bigg(\frac{K-C}{K}+\frac{C}{K}\big[1-\delta\big]^{T_p} \bigg) }_{(D)}= 0. 
\end{align}
\normalsize
When $T_p$ is large enough, $(D)$ is approximated to $ (\frac{K-C}{K}) \ln (\frac{K-C}{K})$. And then, we have 
\par\noindent\small
\begin{align}\label{eq:first_order_simplified}
	 \frac{K-C}{K}  \big( T_p +  r \big) \big[1-\delta \big]^{T_p} \ln (1-\delta) = \frac{K-C}{K} \ln \bigg(\frac{K-C}{K} \bigg).
\end{align}
\normalsize
Note that \eqref{eq:first_order_simplified} has Lambert W-function \cite{corless1996lambertw}, which is defined as when $x e^x = a$, the solution $x$ is $W(a)$, where $W(\cdot)$ is the Lambert W-function. By using the definition of the Lamber W-function, we have the following optimal local iteration $T_p$ from \eqref{eq:first_order_simplified}:
\par\noindent\small
\begin{align}\label{eq:optimal_H_final}
	T_p = \frac{1}{\ln(1-\delta)} W \bigg( \big[1-\delta\big]^r  \ln \bigg( \frac{K-C}{K} \bigg) \bigg) - r.
\end{align}
\normalsize

From the recursive manner of the convergence analysis in a tree network as introduced in Section \ref{sec:analysis}, the optimal number of iterations $T_i$ in Algorithm \ref{alg:general} for a node $Q$ can also be obtained by using aforementioned equation \eqref{prob:primal_iter} with slightly different interpretation. In the tree network, the number of local iterations $T_p$ in  \eqref{prob:primal_iter} is understood as the number of local iteration $T_i$ in Algorithm \ref{alg:general} for the node $Q$. The computational time for the local iteration at a worker, denoted by $t_{lp}$, is interpreted as the computational time for one-time receiving the updating intermediate results from $Q$'s child nodes. And $t_{delay}$ and $t_{cp}$ represent the communication delay time and the processing time at $Q$'s direct parent node respectively. Thus, with the same equation as \eqref{prob:primal_iter} with different interpretation, the optimal number of local iterations for a general tree node $Q$ can be obtained as \eqref{eq:optimal_H_final}.  

\textcolor[rgb]{0,0,0}{Since the objective function $F(T_p)$ in \eqref{prob:primal_iter} represents the convergence bound in terms of time, it is clearly recognized that for a fixed local iteration $T_p$, the larger communication severity $r = t_{delay}/t_{lp}$ exists, the slower convergence rate we have. } Additionally, if in a network, a central node, sub-central nodes and local workers are needed to be chosen, by considering the convergence analysis shown in a recursive manner and the communication delay between layers, choosing a root node making the depth of the connected network shallow will be better for fast convergence.

\textcolor[rgb]{0,0,0}{In the numerical experiments section, we will further investigate the impact of the communication delay severity $r$, and other parameters including $C$, $K$, and $\delta$ in \eqref{eq:optimal_H_final} on the optimal number of local iterations $T_p$.}

\section{Numerical experiments}
\label{sec:experiment}
In wireless communication networks, it can often occur that the local workers are located out of communication range from the central node due to communication constraints such as limited communication power, long distance, limited bandwidth, and limited latency, etc. By reflecting the communication constraints, in the numerical experiments, we consider machine learning scenarios over communication networks, where local workers cannot directly communicate with a central node. Thus, in the distributed dual coordinate ascent for a star network, local workers can only share their local solutions with a central node through multiples of intermediate nodes, which can possibly cause heavy communication delay and latency. For comparison, we solve machine learning problems including regression and classification over different communication networks having different delays with the following datasets: KDD Cup 1998  dataset\footnote{KDD Cup 1998 dataset: \url{https://archive.ics.uci.edu/ml/datasets/KDD+Cup+1998+Data}}, covertype dataset\footnote{Binary Covertype dataset: \url{https://www.csie.ntu.edu.tw/~cjlin/libsvmtools/datasets/binary.html\#covtype.binary}}\cite{blackard1999comparative}, and wine quality dataset\footnote{Wine quality dataset: \url{https://archive.ics.uci.edu/ml/datasets/wine+quality}}\cite{cortez2009modeling}. \textcolor[rgb]{0,0,0}{In addition, we numerically check that the optimal number of local iterations and demonstrate the impact of communication delay on the convergence speed of the distributed dual coordinate ascent by varying the communication delay in networks. And, we further numerically investigate the effect of network topology on the convergence of the distributed dual coordinate ascent over a tree network. }


\textcolor[rgb]{0,0,0}{We compare the convergence of the generalized distributed dual coordinate ascent in tree networks  against that in star networks with intermediate nodes. Since the authors in \cite{jaggi2014communication,ma2015adding} compared the distributed dual coordinate ascent in a star network, so-called CoCoA, with other well known methods including mini-batch SDCA\cite{takac2013mini}, local SGD and mini-batch-SGD\cite{shalev2011pegasos}, we focus on comparing our generalized distributed dual coordinate ascent in tree networks with that in star networks by considering network constraints, especially, communication delay and latency. Additionally, since we are interested in the communication network's effect on the convergence speed of the synchronous distributed dual coordinate ascent, considering the CoCoA+ \cite{ma2015adding}, which is the updated version of CoCoA, or an asynchronous updating method, is out of the scope of this paper.}

\subsection{Machine learning over communication networks}
We consider both regression and classification problems with KDD Cup 1998 dataset and the covtype dataset over communication networks. In the communication networks, we assume that local workers cannot directly reach to a central node, and huge communication delay exists due to the long relay of communication path. In order to reflect this scenario, we deal with various communication delays between the central node and its direct child nodes.


\subsubsection{\textbf{KDD Cup 1998 regression problem}}
\label{subsection:wine} 

In this numerical experiment, we test our algorithm and analysis for a ridge regression problem with KDD Cup 1998 dataset having $481$ attributions including a label and $95412$ instances. We consider the following specific optimization problem by setting $\ell_i(\bw^T \bx_i) = ( \frac{1}{\lambda m} \bw^T \bx_i - y_i)^2$:
\par\noindent\small
\begin{align}\label{eq:wine_primal}
    \underset{\bw \in \R^d}{\text{minimize}}\; \frac{\lambda}{2} ||\bw||^2 + \frac{1}{m} || \bA^T \bw - \by ||^2,
\end{align}
\normalsize
where $\bA \in \R^{d \times m}$ is the feature data matrix whose $i$-th column is $\frac{1}{\lambda m}\bx_i$ and $\by \in \R^m$ is a label vector. Then, the following dual problem is obtained from \eqref{eq:wine_primal}:
\par\noindent\small
\begin{align}
    \underset{\balpha \in \R^m}{\text{maximize}}\; -\frac{\lambda}{2} ||\bA\balpha||^2 - \lambda^2 m \sum_{i=1}^{m} \bigg( \frac{ \alpha_i^2  }{4} - \frac{y_i  \alpha_i}{\lambda m} \bigg).
\end{align}
\normalsize
Hence, in a local worker, $\bigtriangleup \alpha$ in Procedure \ref{alg:leaf} is simply calculated as follows:
\par\noindent\small
\begin{align}
	\bigtriangleup \alpha = - \bigg( \frac{ || \bx_i ||^2}{ \lambda m } + \frac{\lambda^2 m^2}{ 2 } \bigg)^{-1} \bigg( {\bw^{(h-1)}}^T \bx_i + \frac{\lambda^2 m^2}{2} \alpha_i^{(h-1)}  - \lambda m y_i \bigg),
\end{align}
\normalsize
where $(\bx_i, y_i)$ is a randomly chosen data point and $\alpha_i^{(h-1)}$ is $\alpha_i$ value at $(h-1)$-th iteration. 

For the dataset, we take first $95410$ instances and $404$ numerical-type attributions for our numerical experiments. And then, we normalize each attribution with $\ell_2$ norm of it for the performance of regression operation, and then normalize each instance with $\ell_2$ norm in order to make each instance $\bx_i$ hold the condition $|| \bx_i || \leq 1$.   We set the tuning parameter $\lambda$ to $1$. For the communication networks, we consider a tree network model having ten local workers, two sub-central nodes (each having five local workers), and one central node. The simulated star network has ten local workers and one center node. In both cases, we evenly split the data to ten local workers; namely, $9541$ instances without overlap are assigned to each local worker.  

\begin{figure*}[t]
    \centering
   \subfloat[$r=1$]{\includegraphics[scale=0.55]{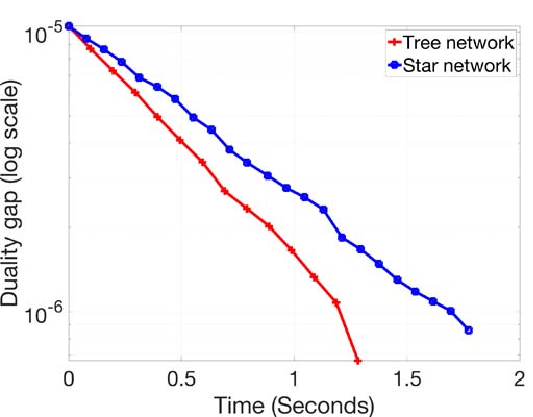}}\;
   \subfloat[$r=10^2$]{\includegraphics[scale=0.55]{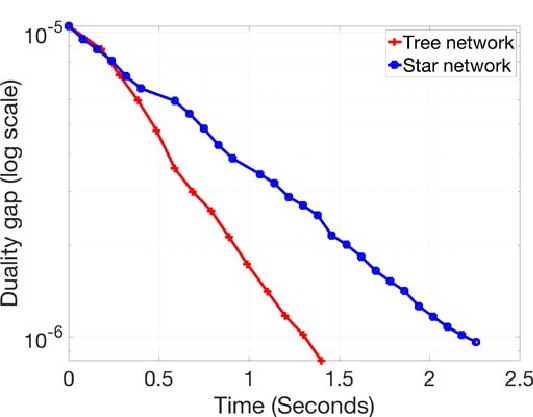}}\;
   \subfloat[$r=10^4$]{\includegraphics[scale=0.55]{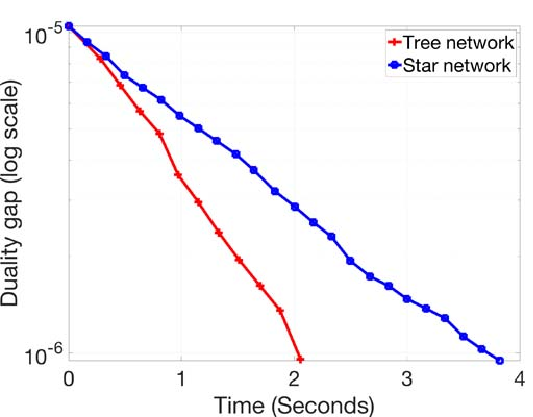}}  
    \caption{ Duality gap at the central node in a regression problem as the operation time of the algorithms goes. The distributed dual coordinate ascent in a tree network (red) and a star network (blue), i.e., CoCoA, are considered when the communication delay, $t_{delay}$, exists between the central node and its direct child nodes. $t_{delay} = r \times t_{lp}$, where $t_{lp}$ represents the computational time for one local iteration at a local worker, and $r$ represents the delay severity level. }
    \label{fig:reg_convergence_reg}
\end{figure*}

We set up a scenario where communication delay, $t_{delay}$, exists between the center node and its direct child node. Therefore, in a star network, the communication delay exists between the central node and local workers, while a tree network has the delay between the central node and the sub-central node. We assume that communication delays between sub-central nodes and local workers are negligible. We set the communication delay $t_{delay} = r \times t_{lp}$, where $t_{lp}$ is the computational time for one local iteration at a worker and the delay severity $r$ is varied from $1$ to $10^4$. Hence, if the delay severity $r$ is huge, then, there exists huge communication delay in the network when it is compared to the local processing time for one iteration. For the algorithm in the tree network, we set the number of local iterations in local workers  and the number of communications between the local workers and the sub-central node to $1000$ and $2$ respectively. For the algorithm in the star network, the number of local iterations at local workers is set to $1000$. \textcolor[rgb]{0,0,0}{Figure \ref{fig:reg_convergence_reg} shows the duality gap at the central node as the operation time goes, and demonstrates that as the communication delay severity increases, the gap between a tree network and a star network in the convergence speed of the distributed algorithm is increased, which indicates  the distributed algorithm in a star network can suffer more from the communication delay effect.}

\subsubsection{\textbf{Covertype dataset classification problem}}
\label{subsubsec:covertype}
We further conduct the comparison between the distributed dual coordinate scent in a star network and a tree network with a standard hinge loss $\ell_2$ regularized SVM. We assume that the communication delay between the central node and its direct child nodes exists in the communcaiton networks. In this experiment, we use the preprocessed Covertype dataset \cite{collobert2002parallel}, which is a binary classification dataset having $581012$ instances and $12$ attributions including label information. The $12$ attributions are expressed as $54$ columns of data with $10$ quantitative variables, $4$ binary wilderness areas and $40$ binary soil type variables.  In order to satisfy the condition $ || \bx_i || \leq 1$, we normalize the dataset and $y_i \in \{ -1, 1\}$, $i=1,...,m$. In this simulation, we organize a tree network having one central node, two sub-central nodes, and eight local workers. Each sub-central node has four local workers. Each local worker has evenly divided instances of the dataset without overlap. For the tree network, the number of communications between the local workers and the sub-central node is set to $10$. The number of local iterations in both networks is set to $300$. 

 \begin{figure*}[t]
    \centering
   \subfloat[$r=1$]{\includegraphics[scale=0.55]{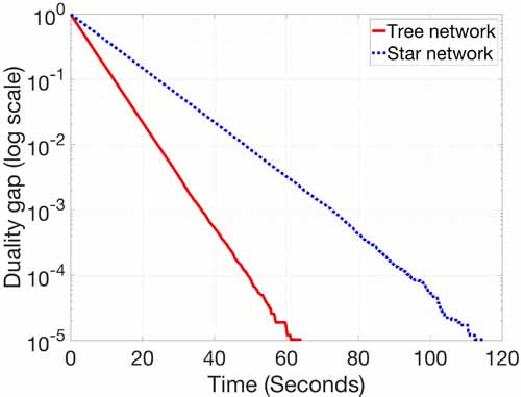}}\;\;
   \subfloat[$r=10^2$]{\includegraphics[scale=0.55]{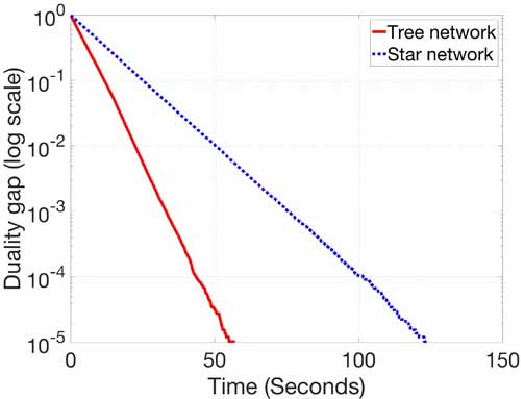}} \;\;
   \subfloat[$r=10^4$]{\includegraphics[scale=0.55]{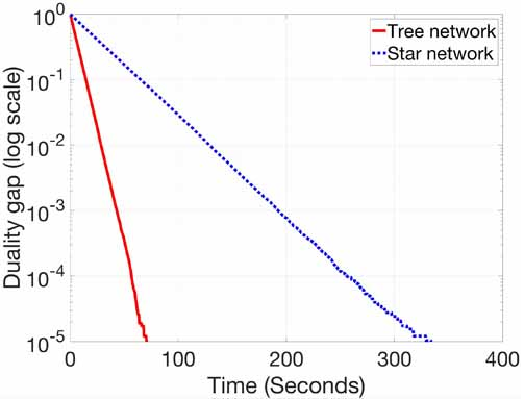}}  
    \caption{Duality gap at the central node in a classification problem as the operation time of the algorithms goes. The distributed dual coordinate ascent in a tree network (red solid line) and a star network (blue dotted line), i.e., CoCoA, are considered when the communication delay, $t_{delay}$, exists between the central node and its direct child nodes. $t_{delay} = r \times t_{lp}$, where $t_{lp}$ represents the computational time for one local iteration at a worker, and $r$ represents the delay severity level.}
    \label{fig:convergence_class}
\end{figure*}

For SVM, we consider the soft-margin SVM classification having hinge loss function, i.e, $\ell_i(\bw^T \bx_i) \triangleq \max( 0, 1 - y_i (\frac{1}{\lambda m}\bw^T \bx_i) )$ as follows:
\par\noindent\small
\begin{align} \label{eq:svm}
    \underset{\bw \in \R^d}{\text{minimize}}\; \frac{\lambda}{2} ||\bw||^2 + \frac{1}{m} \sum_{i=1}^{m} \max( \bm{0},\bm{1} - \bA^T \bw), 
\end{align}
\normalsize
where $\bA_i$, the $i$-th column of the matrix $\bA$, is $\frac{1}{\lambda m}y_i \bx_i$, $\max(\cdot)$ is element-wise operator, and $\bm{0} \in \R^{m}$ and $\bm{1} \in \R^{m}$ are the all $0$ and all $1$ vectors respectively.

Then, the dual problem of \eqref{eq:svm} is stated as follows:
\par\noindent\small
\begin{align}\label{eq:dual_svm}
   & \underset{\balpha \in \R^m}{\text{maximize}}\;  \sum_{i=1}^{m} \alpha_i - \frac{1}{2\lambda} || \bA \balpha ||^2  \quad\quad \text{subject to} \;\;0 \leq \alpha_i \leq \frac{1}{m}, \; \forall i.
\end{align}
\normalsize
Note here that while deriving the dual problem \eqref{eq:dual_svm}, we have $\bw = \frac{1}{\lambda} \bA \balpha$ as the dual-primal variable relation. Then, the local problem for a local worker $Q$ is stated as follows:
\par\noindent\small
\begin{align}\label{eq:dual_local}
   & \underset{\balpha_Q \in \R^{|Q|}}{\text{maximize}}\; - \frac{\lambda}{2} || \overline{\bw} +  \frac{1}{\lambda} \bA_Q \balpha_Q ||^2  +  \sum_{i \in Q} \alpha_i  +  \sum_{i \in \overline{Q}} \alpha_i  \nonumber\\
   & \text{subject to} \;\;0 \leq \alpha_i \leq \frac{1}{m}, \; \forall i \in Q, 
\end{align}
\normalsize
where $\overline{\bw} \triangleq \frac{1}{\lambda} \bA_{\overline{Q}} \balpha_{\overline{Q}} = \bw - \frac{1}{\lambda} \bA_Q \balpha_Q$. Then, in Procedure \ref{alg:leaf} for updating $\bigtriangleup \alpha$, we solve the following optimization problem:
\par\noindent\small
\begin{align}\label{eq:dual_local2}
\bigtriangleup \alpha  = \;\;& \underset{\bigtriangleup \alpha  }{\argmax} - \frac{\lambda}{2} || \bw^{(h-1)} +  \frac{1}{\lambda^2 m} \bigtriangleup \alpha  y_i\bx_i ||^2  +  (\alpha_i^{(h-1)} + \bigtriangleup \alpha )  \nonumber\\
   & \text{subject to} \;\;0 \leq \alpha^{(h-1)}_i + \bigtriangleup \alpha  \leq \frac{1}{m}.
\end{align}
\normalsize
Here, we update the randomly chosen $i$-th coordinate of $\balpha$, where $i \in Q$. It is also possible to update the variable $\balpha_Q$ with a block coordinate method. In order to solve \eqref{eq:dual_local2}, we calculate the optimal solution of \eqref{eq:dual_local2} without the box constraint, i.e., $0 \leq \alpha^{(h-1)}_i + \bigtriangleup \alpha \leq \frac{1}{m}$, and then project the optimal solution onto the box constraint as follows:
\par\noindent\small
\begin{align}
	\bigtriangleup \alpha = \begin{cases} 
											1/m - \alpha^{(h-1)}_i	 \;\; & \text{if}\;\; \alpha^{(h-1)}_i +\bigtriangleup \alpha > 1/m \\
											- \alpha^{(h-1)}_i		\;\;  & \text{if}\;\; \alpha^{(h-1)}_i +	\bigtriangleup \alpha < 0
										\end{cases}.
\end{align} 
\normalsize
Figure \ref{fig:convergence_class} shows the duality gap as the operation time of the algorithms goes. As shown in Figure \ref{fig:convergence_class}, it is better to run more local iterations before sharing intermediate results with the central node when there is huge communication delay in a network.

\subsection{Impact of communication delay on the convergence speed}
\begin{figure}[t]
    \centering
    \subfloat[]{\includegraphics[scale=0.4]{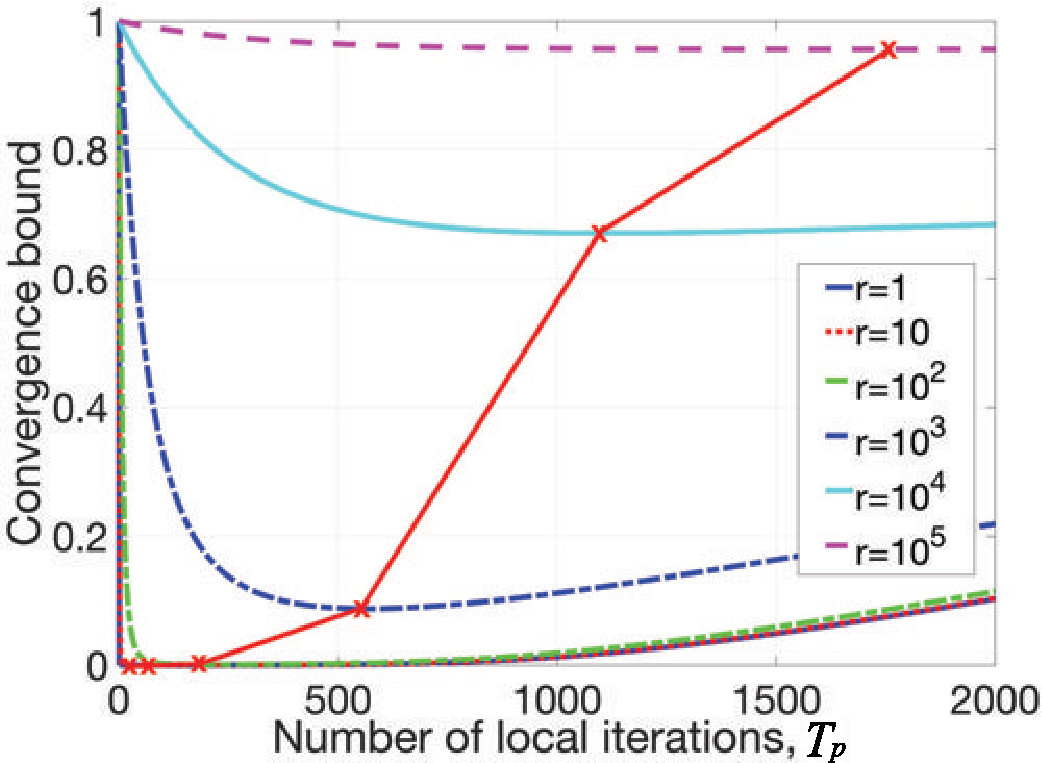}} \quad\quad
    \subfloat[]{\includegraphics[scale=0.78]{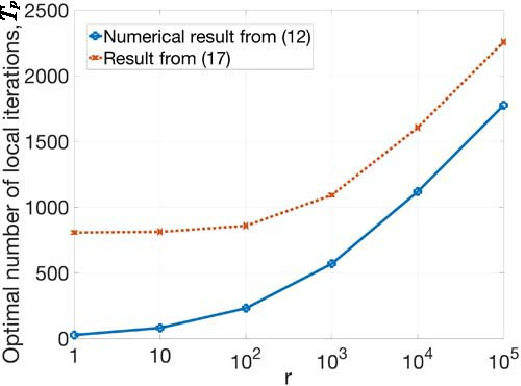}}
    \caption{(a) The objective value of (\ref{prob:primal_iter}), which is the convergence bound (or improvement), when the number of iterations $T_p$ is varied from $1$ to $2000$, where $(C,K,\delta,t_{total},t_{lp}, t_{cp}) = (0.5,3,\nicefrac{1}{300},1,4\times10^{-5},3\times10^{-5})$ and $t_{delay}=r\times t_{lp}$. The red line represents the optimal number of local iterations to achieve the fastest convergence rate. (b) Optimal number of iterations to achieve the fastest convergence rate, when the parameters are the same as (a) and $r$ is varied from $1$ to $10^{5}$. }
    \label{fig:optH}
\end{figure}

In order to see the impact of the communication delay severity $r$, which is the ratio between the communication delay and the local processing time for one iteration, on the optimal number of local iterations $T_p$, we provide Figure \ref{fig:optH} to show the optimal number of local iterations $T_p$ by finding the critical point of \eqref{eq:first_order}. In the simulation, we set $(C,K,\delta,t_{total},t_{lp}, t_{cp}) = (0.5,3,\nicefrac{1}{300},1,4\times10^{-5},3\times10^{-5})$. We set $t_{delay} = r \times t_{lp}$, where $r$ is a parameter indicating how severe the communication delay is. Figure \ref{fig:optH} (a) shows the objective values of (\ref{prob:primal_iter}) when $T_p$ is varied from $1$ to $2000$. The red line represents the optimal convergence bound at the optimal number of local iterations, i.e., the critical point of \eqref{eq:first_order} with different delay severity. Figure \ref{fig:optH} (b) shows the optimal number of local iterations to achieve the fastest convergence rate in different communication delay severity, where $r$ is varied from $1$ to $10^{5}$. The red dotted line is obtained by calculating the given analytical solution introduced in \eqref{eq:optimal_H_final} with given aforementioned parameters, while the blue solid line is obtained by numerically calculating \eqref{prob:primal_iter} and finding the optimal $T_p$ which minimizes the objective value. This simulation results in Figure \ref{fig:optH} show that when the delay severity becomes larger, the more local iterations are desired for the fast convergence speed of the overall algorithm. \textcolor[rgb]{0,0,0}{It is noteworthy that in Figure \ref{fig:optH}(b), the difference between the numerical results from \eqref{prob:primal_iter} and the analytical solution in \eqref{eq:optimal_H_final} is observed. Especially, there is a big gap in the small communication delay severity, e.g., $r=1$. This gap occurs because in the derivation of the analytical solution in \eqref{eq:optimal_H_final}, we approximate (D) of \eqref{eq:first_order2} by assuming that the local iteration $T_p$ is large enough. Hence, the gap becomes smaller when the communication delay severity is increased.}

\begin{figure}[t]
    \centering
    \subfloat[]{\includegraphics[scale=0.8]{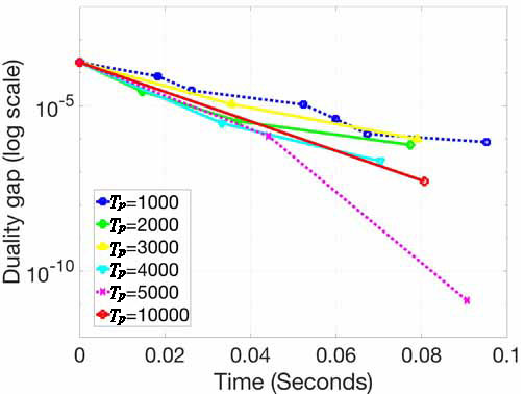}} \quad\quad
    \subfloat[]{\includegraphics[scale=0.8]{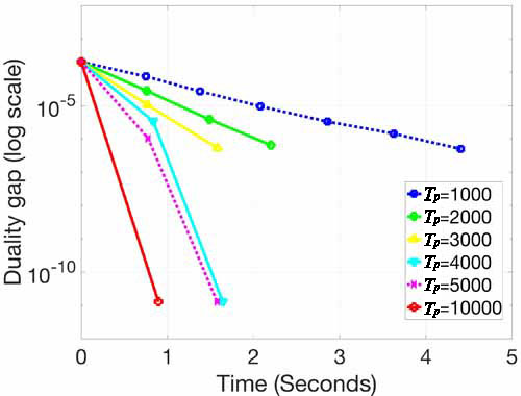}}
    \caption{(a) Duality gap when the delay severity $r$ is $1$. (b) Duality gap when the delay severity $r$ is $10^5$. }
    \label{fig:optH_reg}
\end{figure}

In order to see the impact of the optimal local iterations on a practical machine learning problem, we similarly conduct a regression task with wine quality dataset\cite{cortez2009modeling} in a star network. For the number of iterations in local workers, we vary $T_p$ from $1000$ to $10000$, and evaluate the convergence speed in terms of operation time and duality gap. Figures \ref{fig:optH_reg} (a) and (b) show the duality gap as the operation time goes when the delay severity levels $r$ are set to $1$ and $10^5$ respectively. When $r=1$, the fastest convergence is obtained at $T_p=2000$, while when $r=10^5$, the fastest convergence is obtained at $T_p=10000$. As we expect in Section \ref{sec:optH}, when the communication delay is severe, it is better to perform the more local iterations before sharing the intermediate results with the central node. Also, if the communication delay is small, frequently sharing the intermediate results with the central node is helpful to improve the overall convergence speed. Moreover, we calculate the optimal number of iterations in local workers from the analytical solution \eqref{eq:optimal_H_final} to see whether the analytical solution for the optimal number of local iterations fits to the simulation results. The parameters $\delta$, $K$, and $C$ are set to $\delta=1/1000$, $K=4$, and $C=0.9$ by reflecting the network and simulation settings. With those parameter values, we obtain $2117$ for $r=1$ and $6028$ for $r=10^5$ from the analytical solution in \eqref{eq:optimal_H_final}, while in the simulation, $T_p=2000$ for $r=1$ and $T_p=10000$ for $r=10^5$ provide the best convergence speed. Despite a little difference between the simulation result and the analytical solution for the optimal local number of iterations, \eqref{eq:optimal_H_final} can still be used as a guideline for the number of local iterations in local workers.

\subsection{Network topology's effect on convergence bound}
\begin{figure}[t]
    \centering
  \includegraphics[scale=0.18]{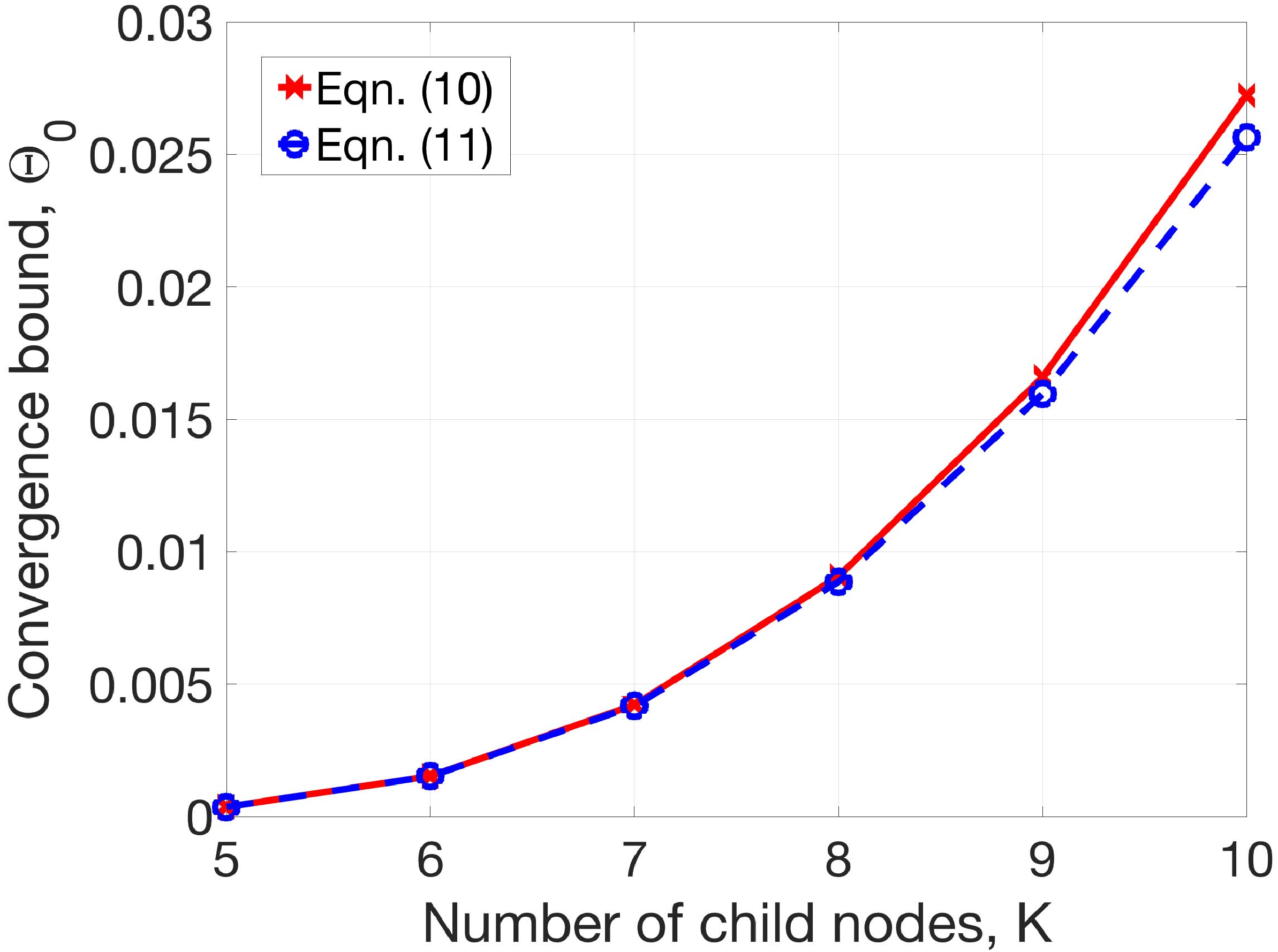}\\
    \caption{Convergence bound over the whole tree network, $\Theta_0$, by varying number of child nodes $K$ with fixed other parameters $(p, T_i, C_i) = (3, 40, 0.9)$ for all $i=0,1,...,p-1$, and $\Theta_p=0.5$. }
    \label{fig:conv_vs_K}
\end{figure}

In this subsection, we numerically investigate the effect of network topology on the convergence bound over the whole tree network, i.e., $\Theta_0$ introduced in \eqref{eq:Theta_0}. In order to see the effect of the number of nodes on the convergence bound, we firstly run simulations by varying the number of child nodes, $K_i$. For the simulation, we take into account a tree network having three layers, i.e., $p=3$. For other parameters, we set $(T_i, C_i)$ to $(40, 0.9)$, for all $i=0,1,...,p$, and $\Theta_p$ to 0.5. We consider that all nodes have the same number of child nodes $K$, i.e., $K_i=K$, for all $i=0,1,...,p-1$, and vary $K$ from 5 to 10. Figure \ref{fig:conv_vs_K} shows the convergence bound $\Theta_0$ by varying the number of child node $K$. The red solid line and the blue dotted line represent the convergence bound expressed in \eqref{eq:Theta_0}, and its approximation introduced in \eqref{eq:Theta_0_approx}. As the number of nodes is increased, the convergence bound $\Theta_0$ is also increased. 

We further run simulations to investigate the effect of the number of layers, $p$, on the convergence bound, $\Theta_0$. For simulations, we set $\frac{K_i-C_i}{K_i} = \frac{K-C}{K}$ and $T_i=T$, for all $i$. From the setting, we can further simplify the approximated convergence bound introduced in \eqref{eq:Theta_0_approx} as 
\par\noindent\small
\begin{align}\label{eq:Theta0_simple}
	\Theta_0 =  & \bigg(\frac{K - C}{K}\bigg)^{T} + \sum_{r=1}^{p-1}\bigg( \frac{K - C}{K}\bigg)^{T(r+1)}  \bigg(\frac{C T}{ K - C} \bigg)^r + \bigg[ \bigg( \frac{K - C }{ K } \bigg)^{T} \bigg( \frac{C T}{K - C} \bigg) \bigg]^p \Theta_p.
\end{align}
\normalsize
For given $K$ and $C$, if $T$ is large enough to be $(\frac{K - C}{K})^{T} \frac{C T}{ K - C}<1$, then,  the convergence bound is expressed as 
 \par\noindent\small
\begin{align}\label{eq:Theta0_largeT}
	\Theta_0 = & \bigg(\frac{K - C}{K}\bigg)^{T} + \bigg( \frac{K - C}{K}\bigg)^{T}  \cdot \frac{ \bigg( \frac{K - C}{K}\bigg)^{T} \frac{C T}{ K - C}  - \bigg( \bigg( \frac{K - C}{K}\bigg)^{T} \frac{C T}{ K - C}  \bigg)^p }{1 - \bigg( \frac{K - C}{K}\bigg)^{T} \frac{C T}{ K - C} } +  \bigg( \frac{K - C }{ K } \bigg)^{T\cdot p} \bigg( \frac{C T}{K - C} \bigg)^p \Theta_p.
\end{align}
\normalsize
Note that $ \lim_{T \rightarrow \infty} (\frac{K - C}{K})^{T} \frac{C T}{ K - C} = 0$.
If $T$ is small enough to be $(\frac{K - C}{K})^{T} \frac{C T}{ K - C} > 1$, then, for the convergence bound, we have
 \par
\begin{figure}[t]
    \centering
 \subfloat[$T=5$]{\includegraphics[scale=0.18]{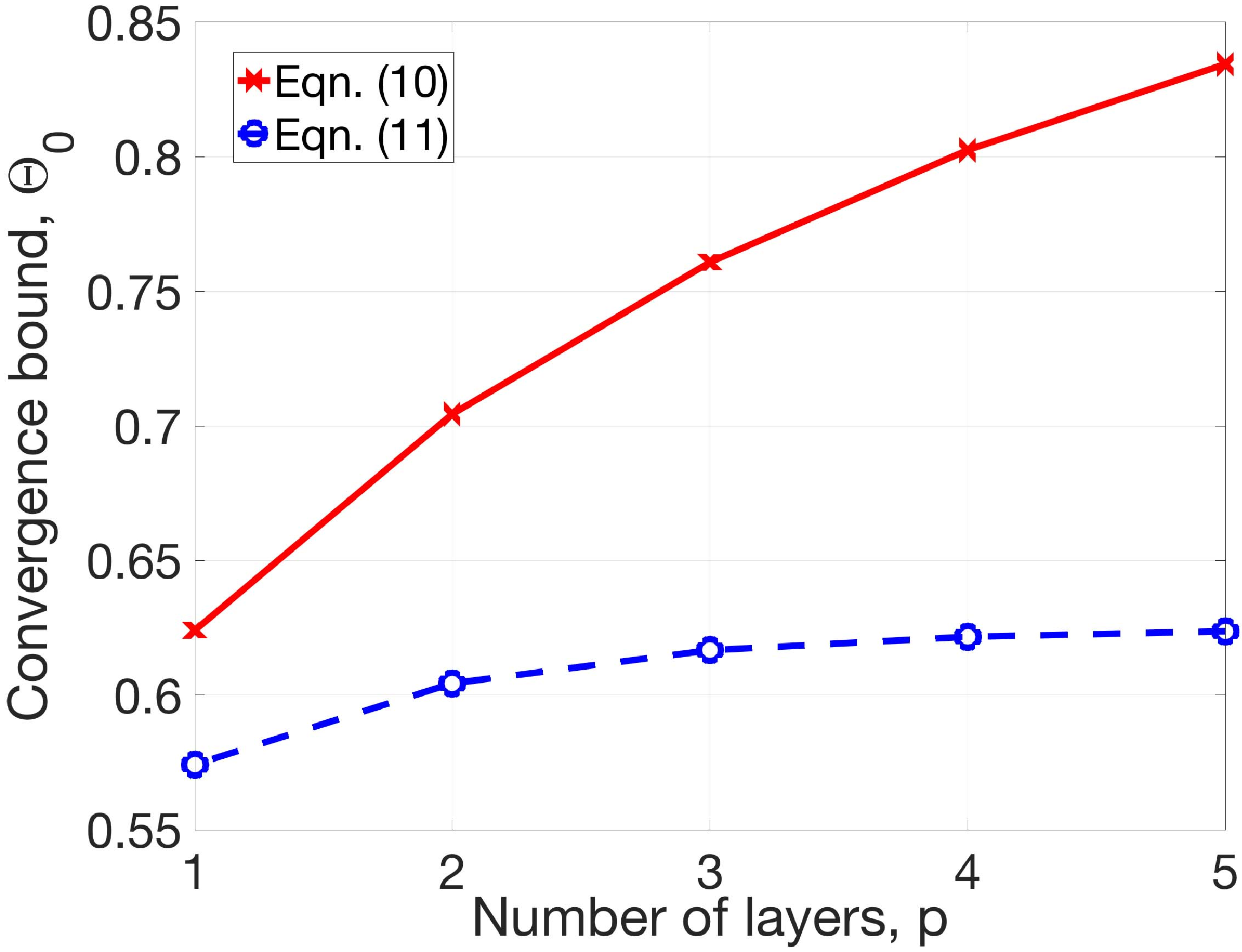}}\;\;
 \subfloat[$T=20$]{ \includegraphics[scale=0.18]{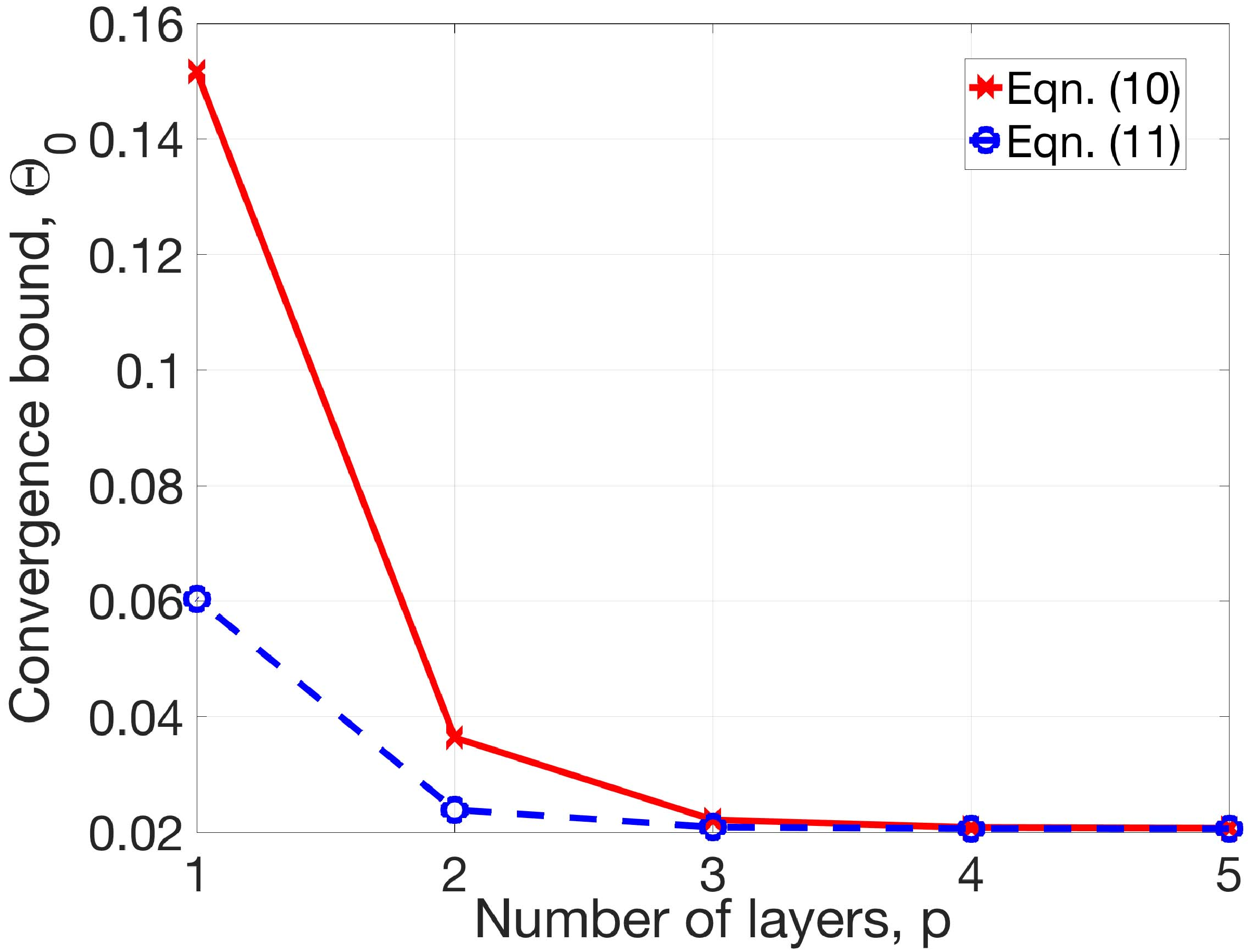}}
    \caption{Convergence bound over the whole tree network, $\Theta_0$, by varying the number of layers $p$ with fixed other parameters $(K, C_i) = (5, 0.9)$ for all $i=0,1,...,p-1$, and $\Theta_p = 0.5$. }
    \label{fig:conv_vs_p} 
\end{figure}

\noindent\small
\begin{align}\label{eq:Theta0_smallT}
	\Theta_0 = & \bigg(\frac{K - C}{K}\bigg)^{T} + \bigg( \frac{K - C}{K}\bigg)^{T}  \cdot \frac{  \bigg( \bigg( \frac{K - C}{K}\bigg)^{T} \frac{C T}{ K - C}  \bigg)^p - \bigg( \frac{K - C}{K}\bigg)^{T} \frac{C T}{ K - C} }{\bigg( \frac{K - C}{K}\bigg)^{T} \frac{C T}{ K - C} - 1} +  \bigg( \frac{K - C }{ K } \bigg)^{T\cdot p} \bigg( \frac{C T}{K - C} \bigg)^p \Theta_p,
\end{align}
\normalsize
Therefore, for given $K$ and $C$, depending on the number of iteration $T$, the dominant term in \eqref{eq:Theta_0} (or \eqref{eq:Theta_0_approx}) is changed like stated in \eqref{eq:Theta0_largeT} or \eqref{eq:Theta0_smallT}, and the convergence bound, $\Theta_0$, follows two different trends as shown in Figure \ref{fig:conv_vs_p}. For Figures \ref{fig:conv_vs_p}(a) and \ref{fig:conv_vs_p}(b), we set the number of iterations $T$ to 5 and 20 respectively with maintaining the other parameters the same. Note that when $p=1$, it represents the star network, and when the number of iteration $T$ is large enough, we can have the better convergence bound in a tree network as shown in Figure 11(b). 

\subsection{Parameter setting for faster convergence speed}

\begin{figure}[t]
    \centering
   \subfloat[$r$ varied]{\includegraphics[scale=0.7]{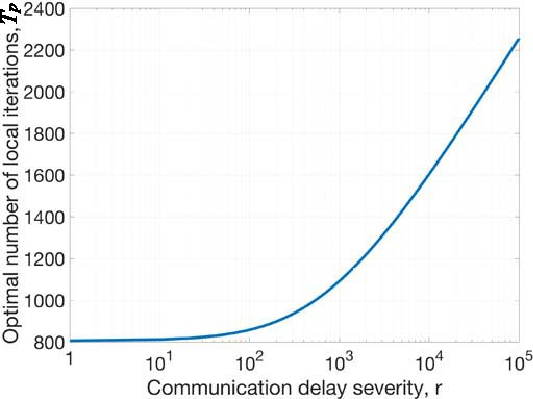}}\quad
   \subfloat[$C$ varied]{\includegraphics[scale=0.7]{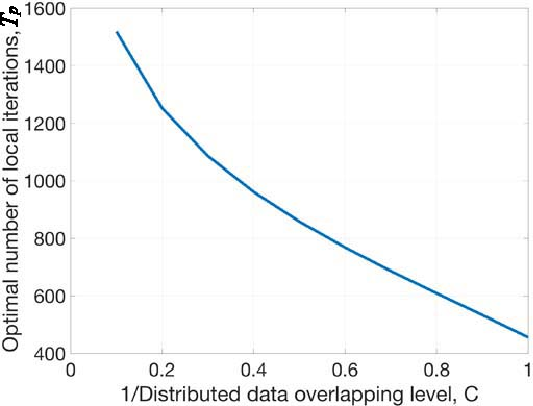}}\\
   \subfloat[$\delta$ varied]{\includegraphics[scale=0.7]{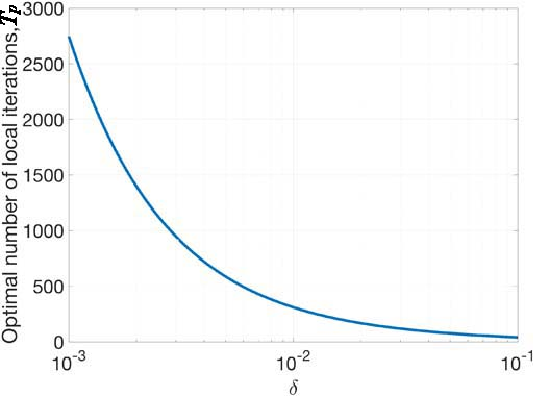}}\quad
   \subfloat[$K$ varied]{\includegraphics[scale=0.7]{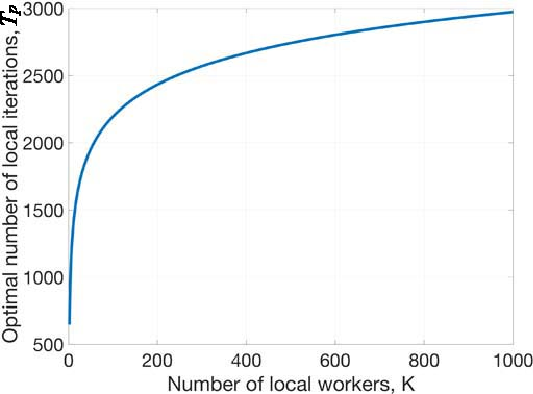}}
    \caption{Optimal number of local iterations, $T_p$, by varying parameters. Except for the varying parameter, other parameters are fixed to $(C,K,\delta, r) = (0.5,3,\nicefrac{1}{300}, 100)$.}
    \label{fig:optimalH_factor}
\end{figure}

\textcolor[rgb]{0,0,0}{In order to investigate the optimal number of local iterations which achieves the fastest convergence speed, from \eqref{eq:optimal_H_final}, we generate Figure \ref{fig:optimalH_factor} by varying each parameter $r$, $C$, $K$, and $\delta$. In Figure \ref{fig:optimalH_factor}(a), the communication delay severity parameter $r$ is varied with fixed other parameters, $(C,K,\delta) = (0.5,3,\nicefrac{1}{300})$. As shown in Figure \ref{fig:optimalH_factor}(a) and the previous subsection, when the communication delay severity $r$ increases, the more number of local iterations before communication with the central node is desired for better convergence rate. Additionally, the parameter $\rho$, which is reciprocal of the parameter $C$ in \eqref{eq:optimal_H_final}, indicates the distributed data overlapping level; namely, smaller $\rho$, less overlapping data among local workers. In order to check the impact of the data overlapping level on the optimal local iteration $T_p$, we vary $C$ with fixing other parameters to $(K,\delta, r) = (3,\nicefrac{1}{300}, 100)$, and draw the graph in Figure \ref{fig:optimalH_factor}(b). From Figure \ref{fig:optimalH_factor}(b), when local workers have more overlapping dataset among them. i.e., larger $\rho$ value or smaller $C$ value, it is desired to run more local iterations to have better convergence speed. And as $\delta$ decreases, correspondingly the step size of the algorithm in a local worker decreases, the more number of local iterations is desired. This is understandable, because with a small step size, more iterations are needed to reach an optimal point. From Figure \ref{fig:optimalH_factor}(d), as the number of local workers, $K$, increases, the optimal number of local iterations, $T_p$, is also increased. Since we fixed other parameters except for $K$, increasing $K$ represents increasing the total size of dataset. And due to the bigger size of dataset in total, we think that more variance in the intermediate results from local workers may lead to more local iterations to reduce the variance.}

\section{Conclusion and discussion} \label{sec:discusion}
In this paper, we study the distributed dual coordinate ascent in a general tree-structured network, where a central node, sub-central nodes and local workers are connected over the communication network, and its analysis. Additionally, since the communication becomes a bottleneck in distributed network systems, we consider the communication delay in time in the convergence analysis of the distributed dual coordinate ascent and obtain the optimal number of iterations to achieve the best convergence speed. \textcolor[rgb]{0,0,0}{In the numerical experiments, we demonstrate the usability of our algorithm and analysis in synchronous machine learning scenarios over communication networks where local workers cannot directly reach to a central node due to communication constraints.}  \textcolor[rgb]{0,0,0}{More specifically, the proposed algorithm in a tree-structured network can reduce the communication overhead at the cost of more local computation complexity. However, since the communication is normally a bottleneck in a distributed process, the distributed algorithm in a tree network can play a significant role in the reduction of communication burden in distributed machine learning process.}


In addition to the work in the paper, the following topics are possible directions for future research. We leave them for the future research. 
\begin{itemize}
\item  \textbf{Asynchronous updating scheme:} Due to the possible performance difference among local workers, it is quite natural to consider asynchronous scheme. Thus, the design and analysis of asynchronous dual coordinate ascent algorithm for generalized tree network topologies can be the next direction of the research.
\item \textbf{Different network topologies:} Since every connected network has its spanning tree, in this paper, a general tree network topology is studied. However, in some network models organized in a mesh, thanks to the network connections in a mesh, the intermediate results from local workers can be easily shared with sub-central nodes and central node or even between local workers. Therefore, the distributed algorithm in mesh networks can have potentials to have faster convergence speed than the algorithm in tree networks. Thus, studying distributed algorithms in mesh networks is of great interest for distributed machine learning operations.
\item \textbf{Various network constraints:} The communication networks can have a variety of network constraints including communication delay, limited communication bandwidth, and limited transmission power.  Motivated by these network constraints, the impact of communication delay on the convergence speed of distributed dual coordinate ascent is studied in this paper. It is also interesting to study the other communication constraints in distributed algorithms.
\item \textbf{Training a neural network over distributed datasets:} Since distributed data can be stored in any communication network, training a neural network over distributed datasets is also of great interest. By considering a spanning tree network, a distributed algorithm framework on a tree-structured network can be a possible approach to this problem. 
\end{itemize}




\ifCLASSOPTIONcaptionsoff
  \newpage
\fi



%

\vspace{-1 em}

%

%
%
%




\appendices
\section{Proof of Theorem \ref{recur_conver}} 
\label{appx:main_proof}
\textcolor[rgb]{0,0,0}{For this proof, we follow the proof of Theorem 2 of \cite{jaggi2014communication} with the additional difference, i.e., dealing with both updating coordinates $\balpha_Q$ and un-updating coordinates $\balpha_{\overline{Q}}$, and show that for a general tree node $Q$, the convergence analysis introduced in \eqref{ieq:convergence} holds.}

\begin{proof}
Suppose the tree node $Q$ has $K$ direct child nodes, and we simply represent the child nodes from $1$ to $K$. The convergence rate of the algorithm at a tree node $Q$ is obtained by considering the updating scheme at the node $Q$ as follows. 
\par\noindent\small
\begin{align}
\label{eq:update}
    \balpha^{(t+1)}=(\balpha^{(t+1)}_{[1:K]},\balpha_{\overline{Q}}) = (\balpha^{(t)}_{[1:K]} + \frac{1}{K} \sum_{k=1}^{K} \bigtriangleup \balpha_{<[k]>},\balpha_{\overline{Q}}),
\end{align}
\normalsize
where $\balpha_{<[k]>}$ is the zero-padding version of $\balpha_{[k]}$ and $Q=[1:K]=\cup_{k=1}^{K}[k]$ is the index set corresponding to workers connected to the node $Q$. The optimal value at the node $Q$ is stated as  
\par\noindent\small
\begin{align}
   D(\balpha_{Q}, \balpha_{\overline{Q}})  
   & = -\frac{\lambda}{2}|| \bA_{Q}\balpha_{Q} + \bA_{\overline{Q}} \balpha_{\overline{Q}} ||^2 - \frac{1}{m}\sum_{i\in Q} \ell_i^{*}(-\alpha_i)  - \frac{1}{m}\sum_{i\in \overline{Q}} \ell_i^{*}(-\alpha_i) \nonumber \\
   & = -\frac{\lambda}{2}||\bA_{[1:K]}\balpha_{[1:K]} +\overline{\bw} ||^2 - \frac{1}{m}\sum_{i\in[1:K]} \ell_i^{*}(-\alpha_i) - \frac{1}{m}\sum_{i\in \overline{Q}} \ell_i^{*}(-\alpha_i), \nonumber
\end{align}
\normalsize
where $\bA_{Q}$ is the partial matrix of $\bA$ by choosing the columns of $\bA$ over the index set $Q$, and $\bA_{\overline{Q}} \balpha_{\overline{Q}}$ is denoted as $\overline{\bw}$.
From (\ref{eq:update}), we have
\par\noindent\small
\begin{align*}
D\big(\balpha^{(t+1)}_{[1:K]}, \balpha_{\overline{Q}}\big) & = D\big(\balpha^{(t)}_{[1:K]} + \frac{1}{K}\sum_{k=1}^{K}\bigtriangleup\balpha_{<[k]>}, \balpha_{\overline{Q}}\big) \\
                            & = D\big(\frac{1}{K} \sum_{k=1}^{K}(\balpha^{(t)}_{[1:K]} + \bigtriangleup\balpha_{<[k]>}), \balpha_{\overline{Q}}\big)\\
                            & \geq \frac{1}{K} \sum_{k=1}^{K} D\big(\balpha^{(t)}_{[1:K]} + \bigtriangleup\balpha_{<[k]>}, \balpha_{\overline{Q}}\big),
\end{align*}
\normalsize
where the inequality is obtained from the Jensen's inequality. Then, we have
\par\noindent\small
\begin{align*}
    D\big(\balpha^{(t+1)}_{[1:K]}, \balpha_{\overline{Q}}\big) - D\big(\balpha^{(t)}_{[1:K]},\balpha_{\overline{Q}}\big) 
    & \geq \frac{1}{K} \sum_{k=1}^{K} \bigg[ D\big(\balpha^{(t)}_{[1:K]} + \bigtriangleup \balpha_{<[k]>},\balpha_{\overline{Q}}\big) - D\big(\balpha^{(t)}_{[1:K]},\balpha_{\overline{Q}}\big)   \bigg]\\
    & = \frac{1}{K} \sum_{k=1}^{K} \bigg[ D\big(\balpha^{(t)}_{[1:K]} + \bigtriangleup \balpha_{<[k]>},\balpha_{\overline{Q}} \big)  \nonumber \\ & \quad\quad - D((\balpha^{(t)}_{[Q;1]}, ..., \balpha^{\star}_{[Q;k]},...,\balpha^{(t)}_{[Q;K]},\balpha_{\overline{Q}} )) \\
    &\quad\quad + D((\balpha^{(t)}_{[Q;1]}, ..., \balpha^{\star}_{[Q;k]},...,\balpha^{(t)}_{[Q;K]},\balpha_{\overline{Q}} )) - D(\balpha^{(t)}_{[1:K]},\balpha_{\overline{Q}})  \bigg]\\
    & = \frac{1}{K} \sum_{k=1}^{K} \bigg[ \epsilon_{Q,k}(\balpha^{(t)}_{[1:K]},\balpha_{\overline{Q}}) - \epsilon_{Q,k}(\balpha^{(t)}_{[1:K]} + \bigtriangleup \balpha_{<[k]>},\balpha_{\overline{Q}}) \bigg],
\end{align*}
\normalsize
where $\epsilon_{Q,k}(\cdot)$ is defined in \eqref{def:epsilon_Qk} and the super-script $\star$ represents the optimal solution. Then, the expectation of $ D\big(\balpha^{(t+1)}_{[1:K]}, \balpha_{\overline{Q}}\big) - D\big(\balpha^{(t)}_{[1:K]},\balpha_{\overline{Q}}\big) $ is lower-bounded as follows:
\par\noindent\small
\begin{align*}
    \E[ D(\balpha^{(t+1)}_{[1:K]},\balpha_{\overline{Q}}) - D(\balpha^{(t)}_{[1:K]},\balpha_{\overline{Q}}) ] 
    & \geq \frac{1}{K} \sum_{k=1}^{K} \bigg[ \E[\epsilon_{Q,k}(\balpha^{(t)}_{[1:K]},\balpha_{\overline{Q}})] - \E[\epsilon_{Q,k}(\balpha^{(t)}_{[1:K]} + \bigtriangleup \balpha_{<[k]>},\balpha_{\overline{Q}} )]    \bigg]\\
    & \geq \frac{1}{K}(1 - \Theta) \sum_{k=1}^{K} \epsilon_{Q,k}(\balpha^{(t)}_{[1:K]},\balpha_{\overline{Q}}),
\end{align*}
\normalsize
where the last inequality is obtained from Assumption \ref{asp:local_assumption}. And $\sum_{k=1}^{K} \epsilon_{Q,k}( \balpha^{(t)}_{[1:K]},\balpha_{\overline{Q}})$ can be bounded as follows.
\par\noindent\small
\begin{align}
    & \sum_{k=1}^{K} \epsilon_{Q,k}( \balpha^{(t)}_{[1:K]},\balpha_{\overline{Q}}) \nonumber\\
    & = \sum_{k=1}^{K} \underset{\hat{\balpha}_{[Q;k]}}{\text{maximize}}\; \bigg[ D((\balpha^{(t)}_{[Q;1]},...,
    \hat{\balpha}_{[Q;k]},...,\balpha^{(t)}_{[Q;K]},\balpha_{\overline{Q}}))  - D((\balpha^{(t)}_{[Q;1]},...,\balpha^{(t)}_{[Q;k]},...,\balpha^{(t)}_{[Q;K]},\balpha_{\overline{Q}})) \bigg] \nonumber\\
    & = \underset{\hat{\balpha} \in \mathbb{R}^{|[1:K]|}}{\text{maximize}}\;
    \sum_{k=1}^{K} \bigg[ D((\balpha^{(t)}_{[Q;1]},...,\hat{\balpha}_{[Q;k]},...,\balpha^{(t)}_{[Q;K]},\balpha_{\overline{Q}})) - D((\balpha^{(t)}_{[Q;1]},...,\balpha^{(t)}_{[Q;k]},...,\balpha^{(t)}_{[Q;K]},\balpha_{\overline{Q}})) \bigg] \nonumber\\
    & = \underset{\hat{\balpha} \in \mathbb{R}^{|[1:K]|}}{\text{maximize}}\;
    \sum_{k=1}^{K} \bigg[ -\frac{\lambda}{2} || \bA_{[1:K]}(\balpha^{(t)}_{[Q;1]},...,\hat{\balpha}_{[Q;k]},...,\balpha^{(t)}_{[Q;K]}) + \overline{\bw}||^2 
    + \frac{\lambda}{2}||\bA_{[1:K]} \balpha^{(t)}_{[1:K]} + \overline{\bw}||^2 \bigg] \nonumber\\
    & \quad\quad\quad\quad\quad\quad - \frac{1}{m} \sum_{i \in [1:K]} \ell^{*}_i(-\hat{\alpha}_i) + \frac{1}{m} \sum_{i\in [1:K]} \ell^{*}_i(-\alpha^{(t)}_i) \nonumber\\
       & = \underset{\hat{\balpha} \in \mathbb{R}^{|[1:K]|}}{\text{maximize}}\;
    \bigg[ \frac{1}{m} \sum_{i \in [1:K]} \bigg( -\ell^{*}_i(-\hat{\alpha}_i) + \ell^{*}_i(-\alpha^{(t)}_i) \bigg) \bigg] -\frac{\lambda}{2}  \sum_{k=1}^{K} \bigg[ || \bA_{[1:K]}(\balpha^{(t)}_{[Q;1]},...,\hat{\balpha}_{[Q;k]},...,\balpha^{(t)}_{[Q;K]}) + \overline{\bw}||^2 \nonumber \\
    & \quad\quad\quad\quad\quad
    - ||\bA_{[1:K]} \balpha^{(t)}_{[1:K]} + \overline{\bw}||^2 \bigg]  \nonumber\\
    & = \underset{\hat{\balpha} \in \mathbb{R}^{|[1:K]|}}{\text{maximize}}\;
    \bigg[ -\frac{1}{m} \sum_{i \in [1:K]} \bigg( \ell^{*}_i(-\hat{\alpha}_i) - \ell^{*}_i(-\alpha^{(t)}_i) \bigg) \bigg]  -\frac{\lambda}{2}  \sum_{k=1}^{K} \bigg[ || \bA_{[1:K]}\balpha^{(t)}_{[1:K]} - \bA_{[k]}(\balpha^{(t)}_{[k]} - \hat{\balpha}_{[k]}) + \overline{\bw}||^2 \nonumber \\
    & \quad\quad\quad\quad\quad
    - ||\bA_{[1:K]} \balpha^{(t)}_{[1:K]} + \overline{\bw}||^2 \bigg] \nonumber \\ 
    & = \underset{\hat{\balpha} \in \mathbb{R}^{|[1:K]|}}{\text{maximize}}\;
    \bigg[ D(\hat{\balpha}_{[1:K]},\balpha_{\overline{Q}})+\frac{\lambda}{2}||\bA_{[1:K]}\hat{\balpha}_{[1:K]} + \overline{\bw} ||^2  - D(\balpha^{(t)}_{[1:K]},\balpha_{\overline{Q}}) - \frac{\lambda}{2} || \bA_{[1:K]}\balpha^{(t)}_{[1:K]} + \overline{\bw}||^2 \bigg] \nonumber\\
    & \quad\quad\quad\quad\quad -\frac{\lambda}{2}  \sum_{k=1}^{K} \bigg[ || \bA_{[1:K]}\balpha^{(t)}_{[1:K]} - \bA_{[k]}(\balpha^{(t)}_{[k]} - \hat{\balpha}_{[k]}) + \overline{\bw}||^2 
  - ||\bA_{[1:K]} \balpha^{(t)}_{[1:K]} + \overline{\bw}||^2 \bigg]  \nonumber\\
   & = \underset{\hat{\balpha} \in \mathbb{R}^{|[1:K]|}}{\text{maximize}}\;
     D(\hat{\balpha}_{[1:K]},\balpha_{\overline{Q}}) - D(\balpha^{(t)}_{[1:K]},\balpha_{\overline{Q}}) +\frac{\lambda}{2}||\bA_{[1:K]}\hat{\balpha}_{[1:K]} + \overline{\bw} ||^2- \frac{\lambda}{2} || \bA_{[1:K]}\balpha^{(t)}_{[1:K]} + \overline{\bw}||^2  \nonumber\\
    & \quad\quad\quad\quad\quad  -\frac{\lambda}{2}  \sum_{k=1}^{K} \bigg[  || \bA_{[k]}(\balpha^{(t)}_{[k]} - \hat{\balpha}_{[k]}) ||^2 - 2(\bA_{[1:K]}\balpha^{(t)}_{[1:K]} + \overline{\bw})^T \bA_{[k]}(\balpha^{(t)}_{[k]} - \hat{\balpha}_{[k]}) \bigg] \nonumber\\
    & = \underset{\hat{\balpha} \in \mathbb{R}^{|[1:K]|}}{\text{maximize}}\;
     D(\hat{\balpha}_{[1:K]},\balpha_{\overline{Q}}) - D(\balpha^{(t)}_{[1:K]},\balpha_{\overline{Q}}) 
    +\frac{\lambda}{2}\bigg(||\bA_{[1:K]}\hat{\balpha}_{[1:K]} + \overline{\bw} ||^2- || \bA_{[1:K]}\balpha^{(t)}_{[1:K]} + \overline{\bw}||^2 \bigg)  \nonumber\\
    & \quad\quad\quad\quad\quad -\frac{\lambda}{2}  \sum_{k=1}^{K} \bigg[|| \bA_{[k]}(\balpha^{(t)}_{[k]} - \hat{\balpha}_{[k]}) ||^2 \bigg] 
+ \lambda (\bA_{[1:K]}\balpha^{(t)}_{[1:K]} + \overline{\bw})^T (\bA_{[1:K]}\balpha^{(t)}_{[1:K]} - \bA_{[1:K]}\hat{\balpha}_{[1:K]} + \overline{\bw} - \overline{\bw})  \nonumber\\
\label{eq:overall} 
    & = \underset{\hat{\balpha} \in \mathbb{R}^{|[1:K]|}}{\text{maximize}}\;
     D(\hat{\balpha}_{[1:K]},\balpha_{\overline{Q}}) - D(\balpha^{(t)}_{[1:K]},\balpha_{\overline{Q}})  +\frac{\lambda}{2}\bigg(||\bA_{[1:K]}\hat{\balpha}_{[1:K]} + \overline{\bw} ||^2- || \bA_{[1:K]}\balpha^{(t)}_{[1:K]} + \overline{\bw}||^2 \bigg)  \nonumber\\
    & \quad\quad\quad\quad -\frac{\lambda}{2}  \sum_{k=1}^{K} \bigg[ || \bA_{[k]}(\balpha^{(t)}_{[k]} - \hat{\balpha}_{[k]}) ||^2 \bigg] + \lambda || \bA_{[1:K]}\balpha^{(t)}_{[1:K]} + \overline{\bw}||^2  - \lambda( \bA_{[1:K]}\balpha^{(t)}_{[1:K]} +\overline{\bw})^T(\bA_{[1:K]}\hat{\balpha}_{[1:K]} + \overline{\bw}) \nonumber \\    
    & = \underset{\hat{\balpha} \in \mathbb{R}^{|[1:K]|}}{\text{maximize}}\;
     D(\hat{\balpha}_{[1:K]},\balpha_{\overline{Q}}) - D(\balpha^{(t)}_{[1:K]},\balpha_{\overline{Q}}) 
     -\frac{\lambda}{2}  \sum_{k=1}^{K} \bigg[ || \bA_{[k]}(\balpha^{(t)}_{[k]} - \hat{\balpha}_{[k]}) ||^2 \bigg] \nonumber\\
     & \quad\quad\quad\quad +\frac{\lambda}{2}\bigg(||\bA_{[1:K]}\hat{\balpha}_{[1:K]} + \overline{\bw} ||^2 + || \bA_{[1:K]}\balpha^{(t)}_{[1:K]} + \overline{\bw}||^2  - 2( \bA_{[1:K]}\balpha^{(t)}_{[1:K]} +\overline{\bw})^T(\bA_{[1:K]}\hat{\balpha}_{[1:K]} + \overline{\bw})  \bigg) \nonumber \\   
    & = \underset{\hat{\balpha} \in \mathbb{R}^{|[1:K]|}}{\text{maximize}}\;
     D(\hat{\balpha}_{[1:K]},\balpha_{\overline{Q}}) - D(\balpha^{(t)}_{[1:K]},\balpha_{\overline{Q}}) 
    -\frac{\lambda}{2} \underbrace{ \bigg[ \sum_{k=1}^{K} \bigg[ || \bA_{[k]}(\balpha^{(t)}_{[k]} - \hat{\balpha}_{[k]}) ||^2 \bigg] - ||\bA_{[1:K]} (\hat{\balpha}_{[1:K]} - \balpha^{(t)}_{[1:K]}) ||^2  \bigg] }_{=(A)} 
\end{align}
\normalsize
We can lower-bound (\ref{eq:overall}) by upper-bounding (A). For the upper-bound of (A), we have
\par\noindent\small
\begin{align*}
	(A) & = \sum_{k=1}^{K} \bigg[ || \bA_{[k]}(\balpha^{(t)}_{[k]} - \hat{\balpha}_{[k]}) ||^2 \bigg] - ||\bA_{[1:K]} (\hat{\balpha}_{[1:K]} - \balpha^{(t)}_{[1:K]}) ||^2   \\
		 & \leq \sum_{i\in [1:K]} ||  \bA_i (\alpha_i^{(t)} - \hat{\alpha}_i ) ||^2 - ||\bA_{[1:K]} (\hat{\balpha}_{[1:K]} - \balpha^{(t)}_{[1:K]}) ||^2 \\		 
		 &  \leq \sum_{i\in [1:K]} \frac{1}{\lambda^2 m^2}  || \bx_i ||^2 (\alpha_i^{(t)} - \hat{\alpha}_i )^2  - ||\bA_{[1:K]} (\hat{\balpha}_{[1:K]} - \balpha^{(t)}_{[1:K]}) ||^2 \\
		 &  \leq \frac{1}{\lambda^2 m^2} \sum_{i\in [1:K]} (\alpha_i^{(t)} - \hat{\alpha}_i )^2  - ||\bA_{[1:K]} (\hat{\balpha}_{[1:K]} - \balpha^{(t)}_{[1:K]}) ||^2 \\
		 &  \leq \frac{1}{\lambda^2 m^2} || \balpha_{[1:K]}^{(t)} - \hat{\balpha}_{[1:K]} ||^2  - ||\bA_{[1:K]} (\hat{\balpha}_{[1:K]} - \balpha^{(t)}_{[1:K]}) ||^2 	\\	 
		 &  \leq \frac{\rho}{\lambda^2 m^2} || \balpha_{[1:K]}^{(t)} - \hat{\balpha}_{[1:K]} ||^2,
\end{align*}
\normalsize
where the second inequality is from  $\bA_i = \frac{1}{\lambda m}  \bx_i$, and  the third inequality is obtained from the assumption of the scaled input data, i.e., $\| \bx_i \| \leq 1$.  We can have the last inequality by introducing $\rho_{min}$, which is the minimum value of $\rho$, to hold the last inequality as follows: 
\par\noindent\small
\begin{align}
    \rho \geq \rho_{min} \triangleq \underset{\balpha \in \R^{| [1:K] |}}{\text{maximize}}\; \lambda^2 m^2\frac{ \sum_{k=1}^{K}  ||\bA_{[k]}\balpha_{[k]}||^2 - ||\bA_{[1:K]} \balpha||^2 }{||\balpha||^2} \geq 0.
\end{align}
\normalsize
The condition $\rho_{min} \geq 0$ can be shown by considering a feasible solution making $\sum_{k=1}^{K}  ||\bA_{[k]}\balpha_{[k]}||^2 - ||\bA_{[1:K]} \balpha||^2  = 0$, e.g., $\balpha = \be_i$, where $\be_i$ is a standard unit vector having 1 in the $i$-th entry and 0 elsewhere. 

Then, (\ref{eq:overall}), which is $\sum_{k=1}^{K} \epsilon_{Q,k}( \balpha^{(t)}_{[1:K]},\balpha_{\overline{Q}})$, is lower-bounded as follows:
\par\noindent\small
\begin{align}
\label{eq:lowerbound_epsilon}
    & \sum_{k=1}^{K} \epsilon_{Q,k}( \balpha^{(t)}_{[1:K]},\balpha_{\overline{Q}}) \nonumber \\
    & \geq  \underset{\hat{\balpha} \in \mathbb{R}^{|[1:K]|}}{\text{maximize}}\;
     D(\hat{\balpha}_{[1:K]},\balpha_{\overline{Q}}) - D(\balpha^{(t)}_{[1:K]},\balpha_{\overline{Q}}) 
    -\frac{\rho}{2 \lambda m^2 }  || \hat{\balpha}_{[1:K]} - \balpha^{(t)}_{[1:K]} ||^2  \nonumber \\
    & \geq  \underset{\eta \in[0,1] }{\text{maximize}}\;
     D\big( \eta\balpha^{\star}_{[1:K]} + (1-\eta) \balpha^{(t)}_{[1:K]},\balpha_{\overline{Q}} \big) - D\big(\balpha^{(t)}_{[1:K]},\balpha_{\overline{Q}}\big)  -\frac{\rho}{2 \lambda m^2 }  || \eta\balpha^{\star}_{[1:K]} + (1-\eta)\balpha^{(t)}_{[1:K]} - \balpha^{(t)}_{[1:K]} ||^2  \nonumber \\
    & \geq  \underset{\eta \in[0,1] }{\text{maximize}}\;
     \eta D\big( \balpha^{\star}_{[1:K]},\balpha_{\overline{Q}} \big) + (1-\eta) D\big(\balpha^{(t)}_{[1:K]},\balpha_{\overline{Q}} \big)  - D\big(\balpha^{(t)}_{[1:K]},\balpha_{\overline{Q}}\big) + \frac{\gamma  \eta(1-\eta)}{2m} ||\balpha^{\star}_{[1:K]} - \balpha^{(t)}_{[1:K]}||^2 \nonumber \\
    & \quad\quad\quad\quad -\frac{\rho \eta^2 }{2 \lambda m^2 }  || \balpha^{\star}_{[1:K]} - \balpha^{(t)}_{[1:K]} ||^2 \nonumber \\   
    & \geq  \underset{\eta \in[0,1] }{\text{maximize}}\;
     \eta D\big( \balpha^{\star}_{[1:K]},\balpha_{\overline{Q}} \big) -\eta D\big(\balpha^{(t)}_{[1:K]},\balpha_{\overline{Q}} \big) + \frac{\gamma\eta(1-\eta)}{2m } ||\balpha^{\star}_{[1:K]}  - \balpha^{(t)}_{[1:K]}||^2 -\frac{\rho \eta^2 }{2 \lambda m^2 }  || \balpha^{\star}_{[1:K]} - \balpha^{(t)}_{[1:K]} ||^2 \nonumber \\
     & =  \underset{\eta \in[0,1] }{\text{maximize}}\;
     \eta D\big( \balpha^{\star}_{[1:K]},\balpha_{\overline{Q}} \big) -\eta D\big(\balpha^{(t)}_{[1:K]},\balpha_{\overline{Q}} \big)  + \frac{\eta}{2m} \bigg( \gamma- \frac{\lambda m \gamma + \rho}{\lambda m} \eta \bigg)   ||\balpha^{\star}_{[1:K]} - \balpha^{(t)}_{[1:K]}||^2,
\end{align}
\normalsize
where $\eta$ in the second inequality is introduced for line search between the optimal solution $\balpha^{\star}_{[1:K]}$ and $\balpha^{(t)}_{[1:K]}$, and the equality holds when $\hat{\balpha}_{[1:K]}$ is in the line between $\balpha^{\star}_{[1:K]}$ and $\balpha^{(t)}_{[1:K]}$. And the third inequality is obtained from the strong concavity of $D(\balpha)$. Specifically, we use the well-known fact that if a function $\ell_i(a)$ is $\frac{1}{\gamma}$-smooth, the conjugate function $\ell_i^{*}$ is $\gamma$ strongly convex: for all $u,v \in \R$ and $\eta \in [0,1]$ \cite{shalev2013stochastic}:
\par\noindent\small
\begin{align}\label{eq:strong-convexity}
	& - \ell_i^{*} (\eta u + (1-\eta) v) \geq - \eta \ell_i^* (u) - (1-\eta) \ell_i^* (v) + \frac{\gamma \eta (1-\eta)}{2} (u - v)^2.
\end{align}
\normalsize
From \eqref{eq:strong-convexity}, we have the following inequality for $D\big( \eta\balpha^{\star}_{[1:K]} + (1-\eta) \balpha^{(t)}_{[1:K]},\balpha_{\overline{Q}} \big)$:
\par\noindent\small
\begin{align*}
	&	D\bigg( \eta\balpha^{\star}_{[1:K]} + (1-\eta) \balpha^{(t)}_{[1:K]},\balpha_{\overline{Q}} \bigg) \nonumber \\
	& = -\frac{1}{2} \bigg\| \bA \big(\eta \balpha^{\star}_{[1:K]} + (1-\eta) \balpha^{(t)}_{[1:K]},\; \balpha_{\overline{Q}}\big) \bigg\|^2  - \frac{1}{m} \sum_{i \in [1:K] } \ell_i^{*} (-\eta \alpha_i^{\star} - (1-\eta)\alpha^{(t)}_i) - \frac{1}{m} \sum_{i \in \overline{Q} } \ell_i^{*} (-\eta \alpha_i - (1-\eta)\alpha_i) \nonumber \\
	& = -\frac{1}{2} \bigg\| \eta \bA \big( \balpha^{\star}_{[1:K]}, \; \balpha_{\overline{Q}}\big) + (1-\eta) \bA \big( \balpha^{(t)}_{[1:K]},\; \balpha_{\overline{Q}}\big) \bigg\|^2 - \frac{1}{m} \sum_{i \in [1:K] } \ell_i^{*} (-\eta \alpha_i^{\star} - (1-\eta)\alpha^{(t)}_i) - \frac{1}{m} \sum_{i \in \overline{Q} } \ell_i^{*} (-\eta \alpha_i - (1-\eta)\alpha_i)  \\
	& \overset{\eqref{eq:strong-convexity}}{\geq} -\frac{1}{2} \bigg\| \eta \bA \big( \balpha^{\star}_{[1:K]}, \; \balpha_{\overline{Q}}\big) + (1-\eta) \bA \big( \balpha^{(t)}_{[1:K]},\; \balpha_{\overline{Q}}\big) \bigg\|^2 - \frac{1}{m} \sum_{i \in [1:K] }  \bigg[ \eta \ell_i^{*} (- \alpha_i^{\star} ) + (1-\eta) \ell_i^{*}(-\alpha^{(t)}_i) - \frac{\gamma \eta(1-\eta)}{2} (\alpha^{\star}_i - \alpha^{(t)}_i)^2 \bigg]  \nonumber \\
    & \quad - \frac{1}{m} \sum_{i \in \overline{Q} } \bigg[ \eta \ell_i^{*} (- \alpha_i ) +  (1-\eta)\ell_i^{*}(-\alpha_i) \bigg] \\
	& \geq -\frac{\eta}{2} \bigg\|  \bA \big( \balpha^{\star}_{[1:K]}, \; \balpha_{\overline{Q}}\big)  \bigg\|^2  - \frac{(1-\eta)}{2} 
	\bigg\|\bA \big( \balpha^{(t)}_{[1:K]},\; \balpha_{\overline{Q}}\big) \bigg\|^2  - \frac{1}{m} \sum_{i \in [1:K] }  \bigg[ \eta \ell_i^{*} (- \alpha_i^{\star} ) + (1-\eta) \ell_i^{*}(-\alpha^{(t)}_i)  - \frac{\gamma \eta(1-\eta)}{2} (\alpha^{\star}_i - \alpha^{(t)}_i)^2 \bigg]  \nonumber \\
    & \quad - \frac{1}{m} \sum_{i \in \overline{Q} } \bigg[ \eta \ell_i^{*} (- \alpha_i ) +  (1-\eta)\ell_i^{*}(-\alpha_i) \bigg]  \\		
	& = -\frac{\eta}{2} \bigg\|  \bA \big( \balpha^{\star}_{[1:K]}, \; \balpha_{\overline{Q}}\big)  \bigg\|^2  - \frac{\eta}{m}  \bigg[\sum_{i \in [1:K] }  \ell_i^{*} (- \alpha_i^{\star} )  + \sum_{i \in \overline{Q} } \ell_i^{*} (- \alpha_i ) \bigg]  - \frac{(1-\eta)}{2} 
	\bigg\|\bA \big( \balpha^{(t)}_{[1:K]},\; \balpha_{\overline{Q}}\big) \bigg\|^2  \nonumber \\
    & \quad - \frac{ (1-\eta) }{m} \bigg[ \sum_{i\in [1:K]} \ell_i^{*}(-\alpha^{(t)}_i) +  \sum_{i \in \overline{Q} } \ell_i^{*} (- \alpha_i)  \bigg] + \frac{\gamma \eta(1-\eta)}{2m} \sum_{i\in [1:K]} (\alpha^{\star}_i - \alpha^{(t)}_i)^2\\
	& = \eta D(\balpha^{\star}_{[1:K]},\balpha_{\overline{Q}}) + (1-\eta) D(\balpha^{(t)}_{[1:K]},\balpha_{\overline{Q}})  + \frac{\gamma \eta(1-\eta)}{2m} \| \balpha^{\star}_{[1:K]} - \balpha^{(t)}_{[1:K]} \|^2.
\end{align*}
\normalsize
Notice that $\eta \in [0,1]$. Also note that we derive the equations by using $\bA ( \balpha^{\star}_{[1:K]}, \balpha_{\overline{Q}} )$; however, at each node, we do not know $\bA_{\overline{Q}}$, but $\overline{\bw}$. Therefore, for the term $\bA ( \balpha^{\star}_{[1:K]}, \balpha_{\overline{Q}} )$,   $(\bA_{Q} \balpha^{\star}_{[1:K]} + \overline{\bw})$ is the correct notation; however in order to clearly show the dual objective function, we use the term $\bA ( \balpha^{\star}_{[1:K]}, \balpha_{\overline{Q}} )$ instead of $(\bA_{Q} \balpha^{\star}_{[1:K]} + \overline{\bw})$ with which the derivation can also go through. 

(\ref{eq:lowerbound_epsilon}) can be lower-bounded by choosing $\eta = \frac{\lambda m \gamma}{\lambda m \gamma + \rho} \geq 0$ as 
\par\noindent\small
\begin{align*}
    (\ref{eq:lowerbound_epsilon}) 
    &  \geq \frac{\lambda m \gamma}{\lambda m \gamma + \rho} \bigg( D\big( \balpha^{\star}_{[1:K]},\balpha_{\overline{Q}} \big) - D\big(\balpha^{(t)}_{[1:K]},\balpha_{\overline{Q}}\big) \bigg) \\
\end{align*}
\normalsize
Therefore, we have
\par\noindent\small
\begin{align}
     \E[ D(\balpha^{(t+1)}_{[1:K]},\balpha_{\overline{Q}}) - D(\balpha^{(t)}_{[1:K]},\balpha_{\overline{Q}}) \;|\;\overline{\bw},\balpha^{(t)}_{[1:K]} ] 
    & \geq \frac{1}{K} (1-\Theta) \sum_{k=1}^K \epsilon_{Q,k}(\balpha^{(t)}_{[1:K]},\balpha_{\overline{Q}}) \nonumber \\
    & \geq \frac{1}{K} (1-\Theta)\frac{\lambda m \gamma}{\lambda m \gamma + \rho} \bigg( D\big( \balpha^{\star}_{[1:K]},\balpha_{\overline{Q}} \big) - D\big(\balpha^{(t)}_{[1:K]},\balpha_{\overline{Q}} \big) \bigg) \label{eq:final_ineq}
\end{align}
\normalsize
From \eqref{eq:final_ineq}, we have
\par\noindent\small
\begin{align*}
    &\E[ D(\balpha^{(t+1)}_{[1:K]},\balpha_{\overline{Q}}) - D(\balpha^{\star}_{[1:K]},\balpha_{\overline{Q}})  + D(\balpha^{\star}_{[1:K]},\balpha_{\overline{Q}}) - D(\balpha^{(t)}_{[1:K]},\balpha_{\overline{Q}}) \;|\;\overline{\bw},\balpha^{(t)}_{[1:K]} ]  \\ 
    & \geq \frac{1}{K} (1-\Theta)\frac{\lambda m \gamma}{\lambda m \gamma + \rho} \bigg( D\big( \balpha^{\star}_{[1:K]},\balpha_{\overline{Q}} \big) - D\big(\balpha^{(t)}_{[1:K]},\balpha_{\overline{Q}} \big) \bigg).
\end{align*}
\normalsize
By moving the term $D(\balpha^{\star}_{[1:K]}) - D(\balpha^{(t)}_{[1:K]},\balpha_{\overline{Q}})$ in LHS to RHS and multiplying $-1$ in both sides, we have 
\par\noindent\small
\begin{align*}
    \E[ D(\balpha^{\star}_{[1:K]},\balpha_{\overline{Q}}) - D(\balpha^{(t+1)}_{[1:K]},\balpha_{\overline{Q}}) \;|\;\balpha^{(t)}_{[1:K]},\overline{\bw} ] 
    & \leq \bigg( 1- \frac{1}{K} (1-\Theta)\frac{\lambda m  \gamma}{\lambda m \gamma + \rho} \bigg) \bigg( D\big( \balpha^{\star}_{[1:K]},\balpha_{\overline{Q}} \big) - D\big(\balpha^{(t)}_{[1:K]},\balpha_{\overline{Q}} \big) \bigg)
\end{align*}
\normalsize
\end{proof}

\section{Derivation of the optimal number of local iterations $T_p$}
For the sake of simplicity of \eqref{eq:first_order}, by denoting $1-\delta$, $\frac{K-C}{K}$, $\frac{C}{K}$, and $(t_{delay} + t_{cp})/ t_{lp}$ to $a$, $b$, $c$, and $r$ respectively, we have the following first order condition over $T_p$ for given $a$, $b$, $c$, and $r$:
\par\noindent\small
\begin{align}\label{eq:first_order_simple}
	b  ( T_p +  r ) a^{T_p} \ln (a) - (b+c a^{T_p} ) \ln (b+ca^{T_p}) = 0,
\end{align}
\normalsize
where $a,b,c \in [0,1)$ and $b+c = 1$. When $T_p$ is large enough, $b(T_p + r)a^{T_p} \ln(a)$ is the dominant term of \eqref{eq:first_order_simple} and notice that $0< a < 1$. Therefore, by approximating the term $(b+ca^{T_p}) \ln ( b+ca^{T_p})$ to $b \ln(b)$, we have $b  ( T_p +  r ) a^{T_p} \ln (a) = b \ln (b)$. And then, the equation is re-stated as follows:
\par\noindent\small
\begin{align*}
	(T_p + r) \ln(a) e^{T_p \ln(a)} =  \ln(b)  
	 & \Rightarrow (T_p + r) \ln(a) e^{(T_p + r) \ln(a) } =  \ln(b) e^{r\ln(a)}
\end{align*}
\normalsize
From the definition of the Lambert W-function, which is when $x e^x = a$, the solution $x$ is $W(a)$, where $W(\cdot)$ is the Lambert W-function, we have 
\par\noindent\small
\begin{align*}
(T_p + r)\ln(a) = W \big( \ln(b) e^{r \ln(a)} \big).
\end{align*}
\normalsize
 Therefore, for the optimal number of local iterations $T_p$, we have 
 \par\noindent\small
\begin{align*}
 T_p = \frac{1}{\ln(a)}W( \ln(b) e^{r \ln(a)} ) - r.
 \end{align*}
\normalsize

\end{document}